\documentclass[12pt]{article}
\usepackage{graphicx}
\usepackage[cp1251]{inputenc}  
\usepackage[russian]{babel}
\usepackage{latexsym}
\usepackage{amssymb}
\usepackage{amsmath}

\begin{document}

\makeatletter
\renewcommand*{\@cite}[2]{{#2}}
\renewcommand*{\@biblabel}[1]{#1.\hfill}
\makeatother

\title{A Three-Dimensional Analytical Model of the Interstellar Extinction within the Nearest Kiloparsec}
\author{\bf \hspace{-1.3cm}\copyright\, 2022 г. \ \ 
G.A.Gontcharov$^1$\thanks{E-mail: georgegontcharov@yahoo.com},
\and \bf A.V.Mosenkov$^{2,1}$,
\and \bf S.S.Savchenko$^{1,3,4}$,
\and \bf V.B.Il'in$^{1,3,5}$,
\and \bf A.A.Marchuk$^{1,3}$,
\and \bf A.A.Smirnov$^{1,3}$,
\and \bf P.A.Usachev$^{1,3,4}$,
\and \bf D.M.Polyakov$^{1,3}$
\and \bf N.Hebdon$^{2}$
}
\date{$^1$ Pulkovo Astronomical Observatory, Russian Academy of Sciences, St. Petersburg, 196140 Russia \\
$^2$ Department of Physics and Astronomy, Brigham Young University, N283 ESC, Provo, UT 84602, USA \\
$^3$ St. Petersburg State University, St. Petersburg, 198504 Russia \\
$^4$ Special Astrophysical Observatory, Russian Academy of Sciences, Nizhnii Arkhyz, Karachai-Cherkessian Republic, 369167 Russia \\
$^5$ St. Petersburg State University of Aerospace Instrumentation, St. Petersburg, 190000 Russia}

\maketitle

\newpage

ABSTRACT
We present a new version of our analytical model of the spatial interstellar extinction variations within the nearest kiloparsec from the Sun. This model treats the three-dimensional 
(3D) dust distribution as a superposition of three overlapping layers: (1) the layer along the Galactic midplane, (2) the layer in the Gould Belt, and (3) the layer passing through 
the Cepheus and Chamaeleon dust cloud complexes. In each layer the dust density decreases exponentially with increasing distance from the midplane of the layer. In addition, 
there are sinusoidal longitudinal extinction variations along the midplane of each layer. We have found the most probable values of 29 parameters of our model using four data sets: 
the 3D stellar reddening maps by Gontcharov (2017), Lallement et al. (2019), and Green et al. (2019) and the extinctions inferred by Anders et al. (2022) for 993291 giants from 
the Gaia Early Data Release 3. All of the data give similar estimates of the model parameters. The extinction for a star or a point in space is predicted by our model with an 
accuracy from $\sigma(A_\mathrm{V})=0.07$ to 0.37 for high and low Galactic latitudes, respectively. The natural fluctuations of the dust medium dominate in these values. 
When ignoring the fluctuations of the medium, the average extinction for an extended object (a galaxy, a star cluster, a dust cloud) or a small region of space is predicted by our 
model with an accuracy from $\sigma(A_\mathrm{V})=0.04$ to 0.15 for high and low Galactic latitudes, respectively. Green et al. (2019) and Anders et al. (2022) give in unison an 
estimate of $A_\mathrm{V}=0.12^m$ for the extinction at high latitudes across the whole Galactic dust half-layer above or below the Sun with the natural fluctuations of the medium
$\sigma(A_\mathrm{V})=0.06^m$. If such a high estimate is subsequently confirmed, then it will require to explain how a substantial amount of dust ended up far from the Galactic midplane. 
Our model is a step in this explanation.
\bigskip\noindent
\leftline {PACS numbers: 98.35.Pr}
\bigskip

\newpage

\section*{INTRODUCTION}

The reddening of a star or the interstellar extinction between the observer and a star is usually determined from its spectral energy distribution based on photometric, astrometric, 
spectroscopic, and other observations. For example, individual extinction estimates for millions of stars have recently been obtained by Anders et al. 
(2022, hereafter AKQ22)\footnote{https://data.aip.de/projects/starhorse2021.html or https://cdsarc.cds.unistra.fr/viz-bin/cat/I/354}
based on photometry and astrometry from the Gaia Early Data Release 3 (EDR3; Gaia 2021a) in combination with photometry from other sky surveys.

Individual reddening/extinction estimates are available or can be obtained not for each object. To provide any object with them, without knowing its characteristics, the dependence 
of the reddening/extinction on Galactic coordinates (longitude $l$, latitude $b$, and heliocentric distance $R$) is analyzed. The estimate for the point in space under consideration 
is actually made based on individual estimates for stars near this point. As a result, three-dimensional (3D) reddening/extinction maps are created in tabular form.

The fluctuations of the interstellar dust medium are typical on a spatial scale of the order of one parsec. As a result of them, the individual reddenings/extinctions for neighboring 
stars can differ significantly (for a discussion, see Green et al. (2015), Green et al. (2019, hereafter GSZ19),\footnote{http://argonaut.skymaps.info/} and Gontcharov (2019)). 
Since the 3D maps use individual reddening/extinction estimates in some spatial window to estimate the reddening/extinction at a point, the maps smooth out the fluctuations of the
medium to a greater or lesser extent, depending on the window size. The size of this window determines the spatial resolution of the map. It is very different for different maps: 
for example, within the nearest kiloparsec from the Sun the GSZ19 map has a spatial resolution $\approx1$ pc, whereas the maps of 
Gontcharov (2017, hereafter G17)\footnote{https://cdsarc.cds.unistra.fr/viz-bin/cat/J/PAZh/43/521} 
and Lallement et al. (2019, hereafter LBV19)\footnote{https://astro.acri-st.fr/gaia\_dev/} 
have a resolution $\approx25$ pc. As a result, the reddening estimates from GSZ19 reflect the fluctuations of the medium, while the estimates from G17 and LBV19 smooth them out strongly.

The reddening/extinction maps with their tabular data presentation do not allow one to uncover the patterns of the large-scale dust distribution in space and to explain the physics 
and geometry of the dust distribution. For this purpose, attempts have been made since the mid-20th century to represent the spatial reddening/extinction variations (or the spatial
dust distribution associated with them) as analytical functions of Galactic coordinates, i.e., as a 3D analytical model of the reddening/extinction (or the dust distribution). 
In particular, the models can give the geometrical characteristics of the Galactic dust layer in the solar neighborhood that match best the observations: the thickness of the layer, 
the dust density in it, and the offset and tilt of the layer with respect to the Galactic midplane. These characteristics are important for understanding the structure
and evolution of our and other galaxies as input data in the construction of Galaxy models and to estimate the characteristics of extragalactic objects at high latitudes that are 
observed through the Galactic dust layer. For example, the spatial dust distribution model by Drimmel et al. (2003) was used in the Besan\c{c}on Galaxy model (Czekaj et al. 2014). In
addition, the reddening/extinction models give initial estimates for a further improvement of both individual reddenings/extinctions (as in AKQ22) and 3D maps (as in GSZ19) by iterations.

The simplest model with an exponential vertical dust distribution in one layer along the Galactic midplane, without any longitudinal variations in the dust distribution, was proposed 
in themid-20th century by Parenago (1954) and other authors. In this model the cumulative extinction A from the observer to a point in space is described by the barometric law:
\begin{equation}
\label{singlelayer}
A=E\,R(1-\mathrm{e}^{-|Z-E_\mathrm{Z}|/E_\mathrm{scale}})E_\mathrm{scale}/|Z-E_\mathrm{Z}|\,,
\end{equation}
where $E$ is the differential extinction in the midplane of the dust layer, $R$ is the distance from the Sun to the point in space, $Z$ is the distance from the point to the
Galactic midplane along the Galactic $Z$ axis,\footnote{The Galactic rectangular coordinate system with the origin in the Sun and the $X$, $Y$, and $Z$ axes directed toward the
Galactic center, in the direction of Galactic rotation, and toward the Galactic north pole, respectively, is considered.} 
$E_\mathrm{Z}$ is the offset of the dust layer midplane with respect to the Galactic midplane along the $Z$ axis, and $E_\mathrm{scale}$ is the scale height of the dust layer.

An alternative 3D analytical model was proposed by Arenou et al. (1992). It approximates the extinction in 199 sky cells by parabolas as a function of heliocentric distance. 
This model was based on the extinction estimates for a small number of stars. However, its main shortcoming was a formal description of the extinction variations without any physical
explanation.

Yet another 3D model of the spatial dust distribution by Drimmel and Spergel (2001) was compared with observational data and was justified by Drimmel et al. (2003). 
This model includes three structural dust components: a warped, but otherwise axisymmetric disk, spiral arms as mapped by known HII regions, and a local Orion-Cygnus arm
segment. Although this model is extended to much of the Galaxy, near the Sun it gives an insufficiently detailed description of the dust distribution, as noted by the authors of the 
model themselves and as shown by Gontcharov and Mosenkov (2017, 2018, 2021b).

The models of Am\^ores and L\'epine (2005) are based on the hypothesis that the extinction is proportional to the gas column density, which, in turn, was taken from HI, CO, and 
dust emission surveys. The first version of their model assumed an axisymmetric Galaxy with dust density variations as a function of Galactic radius and Z coordinate. 
The second version took into account the spiral structure of the Galaxy. Later, their model for an axisymmetric Galaxy was calibrated based on data for a sample of elliptical
galaxies (Am\^ores and L\'epine 2007). The same shortcoming as that of the Drimmel et al. (2003) model is inherent in this model: coverage of a large part of the Galaxy at the 
expense of an insufficiently detailed description of the solar neighborhood.

Being more realistic than the simplest model (1), the later models in one way or another attempted to take into account the well-known distribution of major dust clouds in the sky 
not only along the Galactic equator, but also along the Gould Belt. This distribution was presented, for example, by Dame et al. (2001) in their Fig. 2a.
Dame et al. (2001) noted that ``most of the major local molecular clouds appear to follow Gould's Belt, \ldots the apparent disk of OB stars, gas, and dust surrounding the Sun 
and inclined $\approx20^{\circ}$ to the Galactic plane.''

Gontcharov (2009, hereafter G09) proposed the first 3D analytical model of the dust distribution with the dust layer in the Gould Belt, in addition to the ordinary layer along 
the Galactic midplane. The parameters of this model were derived using several data sets with distance and extinction estimates that should now be deemed inaccurate. The second 
version of this model was presented by Gontcharov (2012b). Its parameters were derived based on the 3D reddening map from Gontcharov (2010) combined with the 3D map of spatial 
variations in the extinction coefficient $R_\mathrm{V}\equiv A_\mathrm{V}/E(B-V)$ from Gontcharov (2012a). Gontcharov (2019) presented the third version of the same model with a 
more realistic geometry of the dust layers. Its parameters were determined using the observed characteristics of a complete sample of red clump giants from the Gaia Data Release 2
(Gaia 2018) in wide Galactic solar neighborhoods. However, our detailed analysis of this sample showed that the dust distribution in the second and fourth Galactic quadrants is 
not fully described by the two dust layers under consideration. In the fourth version of our model (Gontcharov and Mosenkov 2021b) these dust clouds in the second and fourth quadrants
were described as part of the equatorial layer. In the fifth version of our model presented in this paper the same clouds are considered as manifestations of a separate dust layer 
tilted and offset with respect to the equatorial layer and the Gould Belt layer.

A detailed comparison of some reddening/extinction maps and models with one another and with various data sets is presented in our previous studies (Gontcharov 2017; Gontcharov and
Mosenkov 2017a, 2017b, 2018, 2019, 2021a, 2021b). This comparison revealed insurmountable limitations of some models and the corresponding difficulties in creating a model
that is based on up-to-date data and predicts equally accurately the extinction both near the Sun and in much of the Galaxy. In fact, an accurate 3D dust distribution model can now 
be created in the region of the Galaxy where Gaia gives accurate parallaxes and photometry for complete samples of stars of specific classes. Therefore, in this paper, instead of
considering a set of models in a large part of the Galaxy, we made an effort to improve our model using data within the nearest kiloparsec from the Sun. This is also justified for 
the following reason. Ignoring the fluctuations of the medium, the model gives the most accurate predictions at high latitudes, where the fluctuations are minimal. Therefore, the
model is most useful for predicting and analyzing the extinctions of extragalactic objects at high latitudes. Given that at high latitudes almost all of the dust between us and 
extragalactic objects is contained within the nearest kiloparsec, it makes no sense to construct a model beyond it, at least based on present-day data.

It is important that at high latitudes our model is useful far beyond the nearest kiloparsec. Indeed, estimates of the thickness of the Galactic dust layer (see, e.g., Gontcharov 
and Mosenkov 2021b) show significant dust density variations no farther than $|Z|\approx450$ pc. The spatial cylinder under consideration passes through the whole dust layer of such a
thickness at latitudes $|b|>\arctan(450/1000)\approx24^{\circ}$, i.e., for most of the sky. In this part our model can predict the extinction for numerous extragalactic objects.

This paper is organized as follows. In the Section `Model' we describe our model. The data used are described in the Section `Data'. The most probable parameters of our model are 
presented in the Section `Results'. We give remarks to the results in the Section `Discussion'. In the Section `Application to Clusters and Variable Stars' we test the model and
draw conclusions in the Section `Conclusions'.

\section*{MODEL}

As the previous version of our model (Gontcharov and Mosenkov 2021b), the current version represents the extinction to a star or a point in space as a sum of the extinctions in three 
dust layers. Each term is a function of Galactic coordinates. Two terms are the same in both versions (but differ by the values of the parameters): they describe the extinction by a
function that includes a sinusoidal dependence on the Galactic longitude in the equatorial layer and twice the longitude (in the coordinate system of the Gould Belt) in the Belt layer. 
These longitudinal variations in the equatorial layer are justified by the well-known increase in extinction toward the Galactic center and in the Gould Belt by the well-known dust 
concentrations on two opposite sides of the Belt, in the Aquila Rift, Ophiuchus, and Lupus cloud complexes approximately toward the Galactic center and the Taurus, Perseus, Auriga, 
and Orion ones toward the Galactic anticenter. The third term in both versions is introduced to represent the dust concentrations in the second and fourth Galactic quadrants, in Cepheus
($l\approx120^{\circ}$) and Chamaeleon ($l\approx300^{\circ}$), respectively. Consequently, this term describes the extinction by a function that includes a sinusoidal dependence on
twice the longitude, in the equatorial layer (in the previous version) and the separate third layer (in the new version of our model). Below the model parameters referring to the 
layers along the equator, in the Gould Belt, and in the Cepheus-Chamaeleon layer are denoted by the letters `E', `B', and `C', respectively.

The midplanes of the Gould Belt and Cepheus-Chamaeleon layers are tilted to the Galactic midplane by the angles $B_\mathrm{tilt}$ and $C_\mathrm{tilt}$, respectively. Their approximate estimates are: 
$B_\mathrm{tilt}\approx17^{\circ}$ based on the known orientation of the Gould Belt (Perryman 2009, pp. 311-314, 324-328; Bobylev 2014) and 
$C_\mathrm{tilt}\approx-25^{\circ}$ given the latitudes of the Cepheus Flare and Chamaeleon cloud complexes. The positive and negative signs of these estimates mean an orientation of
the layers whereby the sides of these layers with a positive or negative Galactic latitude, respectively, are directed to the Galactic center.

In our model the equatorial layer is treated as an infinite one along the $X$ and $Y$ axes, since it extends far beyond the nearest kiloparsec. The midplane of the equatorial layer 
is offset with respect to the Sun along the $Z$ axis by the distance $E_\mathrm{Z}$.

The midplanes of the other layers are treated as ellipses with the centers offset with respect to the Sun along the $X$, $Y$, and $Z$ axes by the distances
$B_\mathrm{X}$, $B_\mathrm{Y}$, $B_\mathrm{Z}$ and $C_\mathrm{X}$, $C_\mathrm{Y}$, $C_\mathrm{Z}$ for the Gould Belt and the Cepheus-Chamaeleon layer, respectively.
These ellipses have semimajor axes $B_\mathrm{major}$ and $C_\mathrm{major}$, semiminor axes $B_\mathrm{minor}$ and $C_\mathrm{minor}$, and eccentricities
$B_\mathrm{e}$ and $C_\mathrm{e}$. $B_\mathrm{s.m.a.}$ and $C_\mathrm{s.m.a.}$ are the longitudes of the semimajor axes in the coordinate systems of the layers. 
$B_\mathrm{cross}$ and $C_\mathrm{cross}$ are the Galactic longitudes of the lines of intersection between the midplanes of the layers and the Galactic midplane.

The coordinates of a star or a point in space in the coordinate systems of the Gould Belt and the Cepheus-Chamaeleon layer are the latitudes $B_\mathrm{\beta}$ and $C_\mathrm{\beta}$, 
the longitudes $B_\mathrm{\lambda}$ and $C_\mathrm{\lambda}$, and the distances $B_\mathrm{\zeta}$ and $C_\mathrm{\zeta}$ from the midplanes, respectively. 
$B_\mathrm{Rmax}$ and $C_\mathrm{Rmax}$ are the distances from the Sun to the edges of the layers along the line of sight directed to the star/point under consideration. 
We calculate the extinction in each finite layer only either to its edge or to the star/point, depending on what is closer to us. The geometry, tilt, and rotation of the layer in the
Gould Belt are defined by the relations
\begin{equation}
\label{aae}
B_\mathrm{e}^2=1-B_\mathrm{minor}^2/B_\mathrm{major}^2\,,
\end{equation}
\begin{equation}
\label{r0}
B_\mathrm{Rmax}^2=B_\mathrm{minor}^2/(1-[B_\mathrm{e}\cos(B_\mathrm{\lambda}-B_\mathrm{s.m.a.})]^2)\,,
\end{equation}
\begin{equation}
\label{equbeta}
\sin(B_\mathrm{\beta})=\cos(B_\mathrm{tilt})\sin(b)-\sin(B_\mathrm{tilt})\cos(b)\sin(l-B_\mathrm{cross})\,,
\end{equation}
\begin{multline}
\label{equlambda}
\tan(B_\mathrm{\lambda})=[\sin(B_\mathrm{tilt})\sin(b)+\cos(B_\mathrm{tilt})\cos(b)\sin(l-B_\mathrm{cross})]/\cos(b)\cos(l-B_\mathrm{cross})\,,
\end{multline}
\begin{equation}
\label{equzeta}
B_\mathrm{\zeta}=\min(R,B_\mathrm{Rmax})\sin(B_\mathrm{\beta})\,.
\end{equation}
The relations for the Cepheus-Chamaeleon layer are analogous.

As noted previously, the extinction $A$ predicted by our model for a star/point is a sum of three terms, $A=A_E+A_B+A_C$. Each term is analogous to Eq. (1) and depends on the 
longitude along the midplane of the corresponding layer:
\begin{equation}
\label{equ1}
A_E=[E+E_\mathrm{amplitude}\sin(l+E_\mathrm{phase})]R(1-\mathrm{e}^{-|Z-E_\mathrm{Z}|/E_\mathrm{scale}})E_\mathrm{scale}/|Z-E_\mathrm{Z}|\,,
\end{equation}
where $E$, $E_\mathrm{amplitude}$, $E_\mathrm{phase}$ and $E_\mathrm{scale}$ are the constant term, the amplitude, the phase, and the scale height for the equatorial layer. 
The parameters for the other layers are analogous, except for the coefficient $2$ in front of $B_\mathrm{\lambda}$ and $C_\mathrm{\lambda}$ and the factor 
$\min(R,B_\mathrm{Rmax})$ or $\min(R,C_\mathrm{Rmax})$ instead of $R$:
\begin{equation}
\label{equ2}
A_B=[B+B_\mathrm{amplitude}\sin(2B_\mathrm{\lambda}+B_\mathrm{phase})]\min(R,B_\mathrm{Rmax})\cdot(1-\mathrm{e}^{-|B_\mathrm{\zeta}-B_\mathrm{Z}|/B_\mathrm{scale}})B_\mathrm{scale}/|B_\mathrm{\zeta}-B_\mathrm{Z}|\,,
\end{equation}
\begin{equation}
\label{equ3}
A_C=[C+C_\mathrm{amplitude}\sin(2C_\mathrm{\lambda}+C_\mathrm{phase})]\min(R,C_\mathrm{Rmax})\cdot(1-\mathrm{e}^{-|C_\mathrm{\zeta}-C_\mathrm{Z}|/C_\mathrm{scale}})C_\mathrm{scale}/|C_\mathrm{\zeta}-C_\mathrm{Z}|\,.
\end{equation}

\section*{DATA}

Our model treats each point of the interstellar medium as a mutual superposition of the three layers under consideration with fairly large sizes. Consequently, the search for the 
best model parameters using a data set is nondegenerate only when using very accurate distances and extinctions. Moreover, the stars or points of the data set must fill the space 
under consideration sufficiently uniformly. In this case, the accuracy and representativeness of the data far from the Galactic midplane, where the layers are spaced most widely apart, 
are particularly important. Very few data sets with reddening/extinction estimates satisfy these requirements.

The G17 reddening map covers the space within $R<1200$ and $|Z|<600$ pc. This map was obtained by counting millions of main-sequence turnoff stars on the color-magnitude diagram 
from Two Micron All-Sky Survey (2MASS) photometry (Skrutskie et al. 2006). This is the first 3D map based on the distances calibrated from Gaia parallaxes.

The LBV19 extinction map covers the space within $R<3000$ and $|Z|<400$ pc. LBV19 uses Gaia parallaxes in combination with photometry from Gaia and 2MASS to derive the distances and
extinctions by modeling the spatial dust distribution as a Gaussian process. The space density of stars and the accuracy of their parallaxes and photometry drop with increasing 
distance from the Sun. Accordingly, the spatial resolution and accuracy of the LBV19 map drop: only within the nearest kiloparsec do they remain acceptable for our study. 
Moreover, the limitation of the LBV19 map $|Z|<400$ pc makes the estimates of the sought-for parameters of the Gould Belt and Cepheus-Chamaeleon layers inaccurate and biased, because 
LBV19 is poor in data for the parts of these layers farthest from the Galactic midplane.

The GSZ19 map was obtained by applying probabilistic models to the Gaia parallaxes and the Pan-STARRS DR1 (Chambers et al. 2016, hereafter PS1) and 2MASS photometry. 
GSZ19 extends to several kiloparsecs from the Sun, but covers only 3/4 of the sky ($\delta>-30^{\circ}$) due to the use of Pan-STARRS DR1. However, since the Pan-STARRS DR1 stars 
close to the Sun are too bright and photometrically overexposed, the authors of GSZ19 could not estimate the reddening near the Sun closer than some distance that is individual for 
each line of sight. This distance is typically $250-300$ pc. The absence of GSZ19 estimates for a quarter of the sky and within several hundred parsecs from the Sun must bias 
significantly the estimates of the sought-for parameters of our model. To suppress these biases, we limited the space where the GSZ19 estimates are considered by distance $(X^2+Y^2)^{0.5}$.

Thus, the mentioned features forced us to limit the space under consideration by the cylinder within $(X^2+Y^2)^{0.5}<1000$ and $|Z|<1000$ pc identically for all maps. 
This cylinder goes slightly beyond the nearest kiloparsec.

The original reddening/extinction estimates from G17, LBV19, and GSZ19 were converted to the estimates of the extinction $A_\mathrm{V}$ in the $V$ band using the extinction law from 
Cardelli et al. (1989, hereafter CCM89) with $R_\mathrm{V}=3.1$.

The G17, LBV19, and GSZ19 maps allow the extinction to be interpolated for an infinite set of lines of sight. Therefore, for each map we chose randomly 256\,000 points uniformly 
distributed in the space under consideration in which the extinction was used to determine the model parameters.

Apart from the three reddening/extinction maps, we used the AKQ22 data set with individual $A_\mathrm{V}$ estimates for a set of stars. Apparently, these are the most accurate 
extinction estimates based on Gaia EDR3 results. Therefore, in our study AKQ22 is particularly important as the prototype of future Gaia results. The declared accuracy of the 
$A_\mathrm{V}$ estimates from AKQ22 is $0.13^m$ and $0.15^m$ for stars with a magnitude in the broad Gaia filter $G=14^m$ and $17^m$, respectively.

AKQ22 does not have the limitations noted by us for the three maps. This allowed the space with $(X^2+Y^2)^{0.5}>1000$ pc to be also considered. It turned out that our model is 
approximately equally successful in predicting the extinction up to $(X^2+Y^2)^{0.5}=1200$ pc, i.e., it gives an approximately constant mean `AKQ22 minus best model prediction'
standard deviation and an approximately constant correlation coefficient. Near the Galactic midplane at $(X^2+Y^2)^{0.5}>1200$ pc the correspondence of the model to the AKQ22 
estimates worsens due to the distant dust containers outside the three-layer geometry being considered by us (see the Section `Discussion').

To determine the parameters of our model, we used various samples of stars from AKQ22. Using only giants gave more reliable results (least dependent on the composition and spatial 
distribution of the sample) than using only stars of the main sequence or all classes.

The final sample of the best data from AKQ22 includes 993\,291 stars within $(X^2+Y^2)^{0.5}<1200$ and $|Z|<1000$ pc with a relative accuracy of the distances better than 10 per cent, 
an uncertainty in $A_\mathrm{V}$ better than $0.2^m$, an absolute magnitude $0<M_G<3.3$ (this criterion selects giants), a color in the range $0.85<(BP-RP)_0<1.71$ using the Gaia
$BP$ and $RP$ filters (this criterion refines the selection of giants), \verb"fidelity" $>0.5$, \verb"output flag =%%00",
and a \verb"phot_bp_rp_colour_excess" limitation in accordance with Eq. (18) from Gaia (2021b). The parameters of our model from its comparison with the complete sample of 993\,291 
AKQ22 giants and the randomly selected subsamples of 256\,000 giants turned out to be approximately the same.

Note that 24\,904 (2.5 per cent) among the selected 993\,291 AKQ22 giants have negative extinction estimates.

\begin{table*}
 \centering
\def\baselinestretch{1}\normalsize\normalsize
\caption[]{The parameters of our model derived for the data sets under consideration. The values of the parameters that were not used in calculating the means are given in parentheses. 
The values of the parameters dependent on other parameters are italicized. 
}
\label{solution}
\begin{tabular}[c]{lccccc}
\hline
\noalign{\smallskip}
Parameter                             & G17 & LBV19 & GSZ19 & AKQ22 & Mean \\
\hline
\noalign{\smallskip}
\multicolumn{6}{c}{Equatorial layer} \\
\noalign{\smallskip}
$E_\mathrm{Z}$ (pc)                        & $15$      & $29$     & $20$      & $30$     & $24\pm7$ \\ 
$E$ (mag kpc$^{-1}$)                       & $0.88$    & $0.58$   & $1.05$    & $0.88$   & $0.85\pm0.20$ \\ 
$E_\mathrm{amplitude}$ (mag kpc$^{-1}$)    & $0.54$    & $0.58$   & $1.05$    & $0.55$   & $0.68\pm0.25$ \\ 
$E_\mathrm{phase}$ (deg)                   & $42$      & $23$     & $41$      & $28$     & $33\pm10$ \\ 
$E_\mathrm{scale}$ (pc)                    & $65$      & $40$     & $38$      & $58$     & $50\pm13$ \\ 
\multicolumn{6}{c}{Gould Belt layer} \\
\noalign{\smallskip}
$B_\mathrm{X}$ (pc)                        & $9$       & $0$      & $(32)$    & $5$      & $5\pm5$ \\ 
$B_\mathrm{Y}$ (pc)                        & $68$      & $90$     & $(30)$    & $68$     & $75\pm13$ \\
$B_\mathrm{Z}$ (pc)                        & $6$       & $-25$    & $(14)$    & $9$      & $-3\pm19$ \\
$B_\mathrm{tilt}$ (deg)                    & $17$      & $10$     & $15$      & $16$     & $15\pm3$ \\
$B_\mathrm{cross}$ (deg)                   & $-89$     & $-92$    & $-94$     & $-90$    & $-91\pm2$ \\
$B_\mathrm{s.m.a.}$ (deg)                  & $92$      & $102$    & $100$     & $95$     & $97\pm5$ \\
$B$ (mag kpc$^{-1}$)                       & $1.18$    & $0.95$   & $1.03$    & $1.02$   & $1.05\pm0.10$ \\
$B_\mathrm{amplitude}$ (mag kpc$^{-1}$)    & $1.18$    & $0.94$   & $1.03$    & $1.02$   & $1.04\pm0.10$ \\
$B_\mathrm{phase}$ (deg)                   & $-84$     & $-99$    & $-104$    & $-90$    & $-94\pm9$ \\
$B_\mathrm{scale}$ (pc)                    & $67$      & $58$     & $50$      & $62$     & $59\pm7$ \\
$B_\mathrm{minor}$ (pc)                    & $136$     & $154$    & $220$     & $155$    & $166\pm37$ \\
$B_\mathrm{major}$ (pc)                    & $690$     & $730$    & $830$     & $720$    & $743\pm61$ \\
$B_\mathrm{e}$                             & $\mathit{0.979}$ & $\mathit{0.977}$ & $\mathit{0.964}$ & $\mathit{0.977}$ & $\mathit{0.975\pm0.007}$ \\
Offset of Gould Belt center (pc)           & $\mathit{69}$    & $\mathit{93}$    & $(\mathit{46})$    & $\mathit{69}$    & $\mathit{76\pm24}$ \\
\multicolumn{6}{c}{Cepheus-Chamaeleon layer} \\
\noalign{\smallskip}
$C_\mathrm{X}$ (pc)                        & $-75$     & $-80$    & $(-40)$   & $-74$    & $-76\pm3$ \\
$C_\mathrm{Y}$ (pc)                        & $5$       & $-10$    & $(0)$     & $0$      & $-2\pm8$ \\
$C_\mathrm{Z}$ (pc)                        & $-5$      & $0$      & $(48)$    & $0$      & $-2\pm3$ \\
$C_\mathrm{tilt}$ (deg)                    & $-42$     & $-18$    & $(-4)$    & $-23$    & $-28\pm13$ \\
$C_\mathrm{cross}$ (deg)                   & $-80$     & $-86$    & $(-87)$   & $-91$    & $-86\pm6$ \\ 
$C_\mathrm{s.m.a.}$ (deg)                  & $30$      & $29$     & $(29)$    & $23$     & $27\pm4$ \\
$C$ (mag kpc$^{-1}$)                       & $0.85$    & $0.47$   & $(0.65)$  & $0.38$   & $0.57\pm0.25$ \\
$C_\mathrm{amplitude}$ (mag kpc$^{-1}$)    & $0.24$    & $0.47$   & $(0.65)$  & $0.38$   & $0.36\pm0.12$ \\
$C_\mathrm{phase}$ (deg)                   & $132$     & $62$     & $(25)$    & $47$     & $80\pm45$ \\
$C_\mathrm{scale}$ (pc)                    & $155$     & $60$     & $(48)$    & $90$     & $102\pm49$ \\
$C_\mathrm{minor}$ (pc)                    & $230$     & $180$    & $(266)$   & $250$    & $220\pm36$ \\
$C_\mathrm{major}$ (pc)                    & $1000$    & $1000$   & $(1000)$  & $1200$   & $>1000$ \\
$C_\mathrm{e}$                             & $\mathit{0.973}$  & $\mathit{0.984}$  & $\mathit{(0.964)}$ & $\mathit{0.978}$  & $\mathit{0.978\pm0.005}$ \\
Offset of Cepheus layer center (pc)        & $\mathit{75}$     & $\mathit{81}$     & $(\mathit{62})$      & $\mathit{74}$     & $\mathit{76\pm9}$ \\
\hline
\noalign{\smallskip}
Standard deviation of $A_\mathrm{V}$ residuals        & $0.17$    & $0.24$   & $0.33$   & $0.35$   &  \\
Correlation coefficient                               & $0.86$    & $0.83$   & $0.77$   & $0.73$   & \\ 
\hline
\end{tabular}
\end{table*}


\section*{RESULTS}

The new version of our model has 29 sought-for parameters: 5, 12, and 12 for the layer along the Galactic midplane, in the Gould Belt, and in the Cepheus-Chamaeleon layer, 
respectively. When calculating the most probable values of the parameters, each parameter was varied with a sufficiently small step in a sufficiently wide range. For example, we
varied all of the angular parameters in the entire range of angles $0^{\circ}-360^{\circ}$ with a $0.5^{\circ}$ step. Present-day computational resources allowed us to consider several
trillion sets of parameters in their 29-dimensional space. For each set of parameters we calculated the `data set minus model' residuals, their mean and standard deviation, and the 
linear correlation coefficient between the model predictions and the estimates from the data set. The set of parameters that gives zero mean residual, minimum standard deviation,
and maximum correlation coefficient is the most probable one (i.e., the sought-for solution).

Table 1 presents the most probable values of the parameters found by us for each data set. Note that, for convenience, Table 1 gives our estimates of the semiminor and semimajor 
axes and the eccentricities, although only two of these three parameters are independent. Table 1 gives not only the offsets of the layers with respect to the Sun along the coordinate
axes, but also the total offsets $(B_\mathrm{X}^2+B_\mathrm{Y}^2+B_\mathrm{Z}^2)^{0.5}$ and $(C_\mathrm{X}^2+C_\mathrm{Y}^2+C_\mathrm{Z}^2)^{0.5}$ for the Gould Belt and 
Cepheus-Chamaeleon layers, respectively.

The values of the parameters found in Table 1 are consistent with the constraints that can be imposed on the parameters based on the known locations of the largest dust clouds and 
other estimates of the parameters from the literature: Dame et al. (2001), Kun et al. (2008), Perryman (2009), Gontcharov (2012c), Chen et al. (2020), Spilker et al. (2021) and 
references therein. More specifically, the following is worth expecting:
\begin{itemize}
\item $10<E_\mathrm{Z}<30$ pc,
\item $35<E_\mathrm{scale}<200$ pc,
\item $20^{\circ}<E_\mathrm{phase}<60^{\circ}$, implying the tendency for the extinction to increase from the third Galactic quadrant to the first one,
\item $B_\mathrm{tilt}\approx18^{\circ}$,
\item $B_\mathrm{cross}\approx-90^{\circ}$, implying that the Gould Belt rises above and sinks below the Galactic equator approximately toward the Galactic center and anticenter, 
respectively,
\item $B_\mathrm{s.m.a.}\approx+90^{\circ}$, implying that the semi-major axis of the Gould Belt is directed approximately along the $X$ axis,
\item $B_\mathrm{phase}\approx-90^{\circ}$, implying the maximum extinction in the Gould Belt layer along its semi-major axis,
\item $B_\mathrm{major}<1000$ pc as an estimate of the Gould Belt size,
\item $C_\mathrm{major}\ge1000$ pc as an estimate of the distance to most of the dust clouds in Cepheus (Cepheus Flare dust complex),
\item the shift of the Gould Belt center $50<(B_\mathrm{X}^2+B_\mathrm{Y}^2+B_\mathrm{Z}^2)^{0.5}<150$ pc.
\end{itemize}

The low standard deviations and the high correlation coefficients in Table 1 imply that our model is a good approximation of the data sets under consideration.
Table 1 shows that the data sets give similar values of the parameters. This allows us to consider the values of the parameters averaged for the four data sets and their standard 
deviations that are presented in the right column of Table 1 (as noted previously, the estimates of some parameters can be strongly biased - in Table 1 they are taken in parentheses 
and were not used in the averaging). These standard deviations seem to be a more informative estimate of the accuracy of the derived parameters than other estimates.

The characteristics in Table 1 can be divided into the dust density characteristics 
$E$, $E_\mathrm{amplitude}$, $E_\mathrm{phase}$, $E_\mathrm{scale}$, $B$, $B_\mathrm{amplitude}$, $B_\mathrm{phase}$, $B_\mathrm{scale}$, 
$C$, $C_\mathrm{amplitude}$, $C_\mathrm{phase}$, $C_\mathrm{scale}$ 
and the remaining, geometrical characteristics of the layers describing their size, shape, orientation, and offset with respect to the Sun. The former differ noticeably for different 
data sets, while the latter are fairly close (probably, with the exception of the tilt $C_\mathrm{tilt}$ of the Cepheus-Chamaeleon layer, which, hopefully, will be determined more 
accurately with the appearance of more accurate data sets). The difference in the dust density characteristics is due to the data errors, the difference in the spatial and angular 
resolutions of the data sets, the difference in the data set creation methods, and the possible spatial variations of the extinction law.
These variations can be important, since we convert the original estimates of the data sets obtained in different wavelength ranges to $A_\mathrm{V}$ using the CCM89 extinction law 
with $R_\mathrm{V}=3.1$. More specifically, we use the 2MASS reddening $E(J-Ks)$ from G17, the reddening $E(g_{PS1}-r_{PS1})$ from GSZ19, the extinction $A_\mathrm{0}$ at a wavelength 
of 550 nm from LBV19, and the extinction $A_\mathrm{V}$ from AKQ22, actually calculated by the authors from $A_\mathrm{G}$. In future, with the appearance of more extensive and accurate 
data sets and/or when revealing significant systematic errors in some data set, our model can take into account the spatial variations of the extinction law and/or be scaled by 
including the errors found. In this case, the geometrical characteristics of the layers will change insignificantly, if at all, but will be significantly improved and, hopefully, 
only the dust density characteristics will converge to some unified values.

\begin{figure*}
\includegraphics{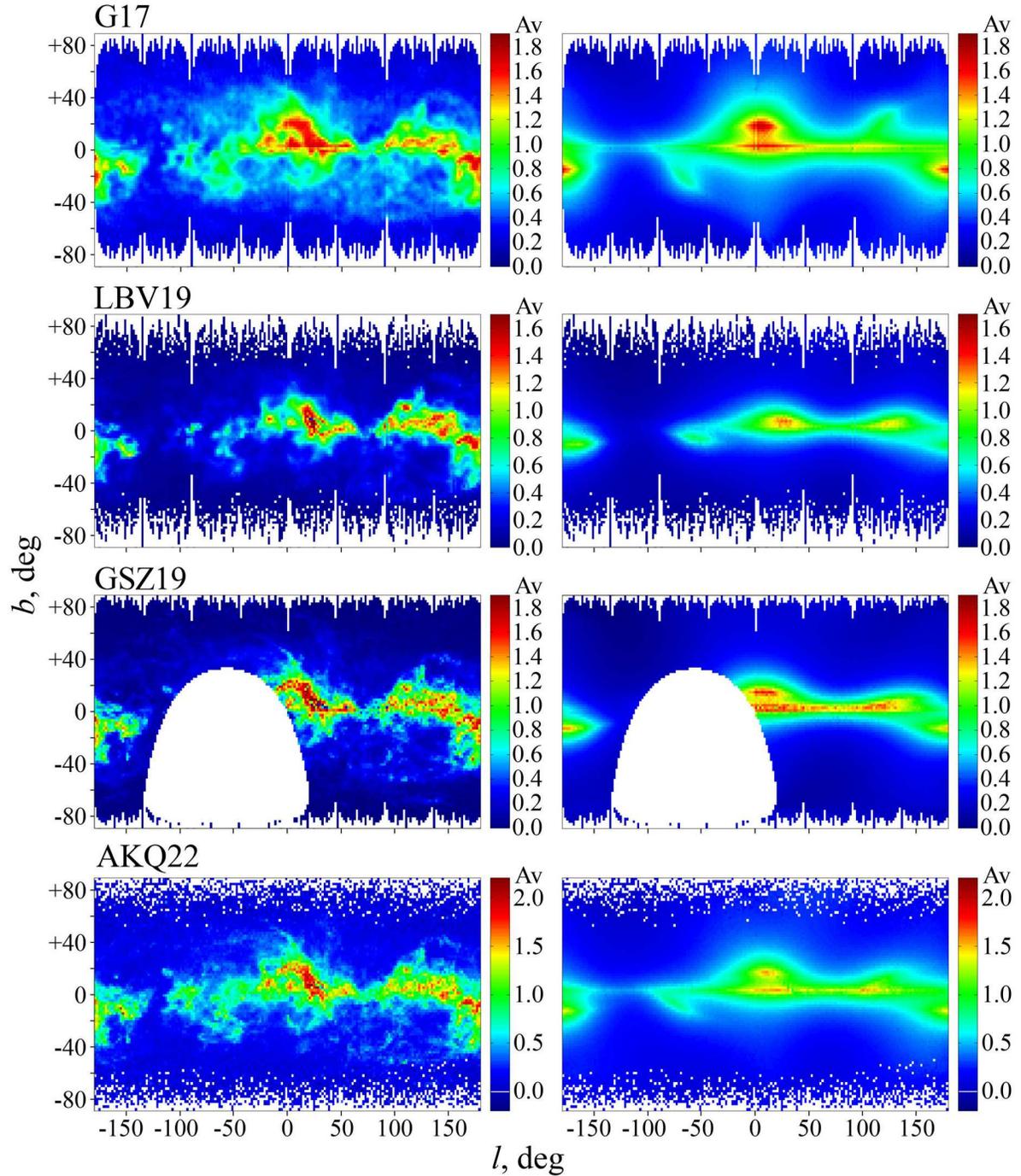}
\caption{Extinction $A_\mathrm{V}$ averaged in $2\times2$ deg$^2$ cells as a function of Galactic coordinates: the estimates from various data sets (the left column of graphs) and 
the best model predictions (the right column of graphs). $A_\mathrm{V}$ is represented by the color scale on the right.
}
\label{oclb}
\end{figure*}

\begin{figure*}
\includegraphics{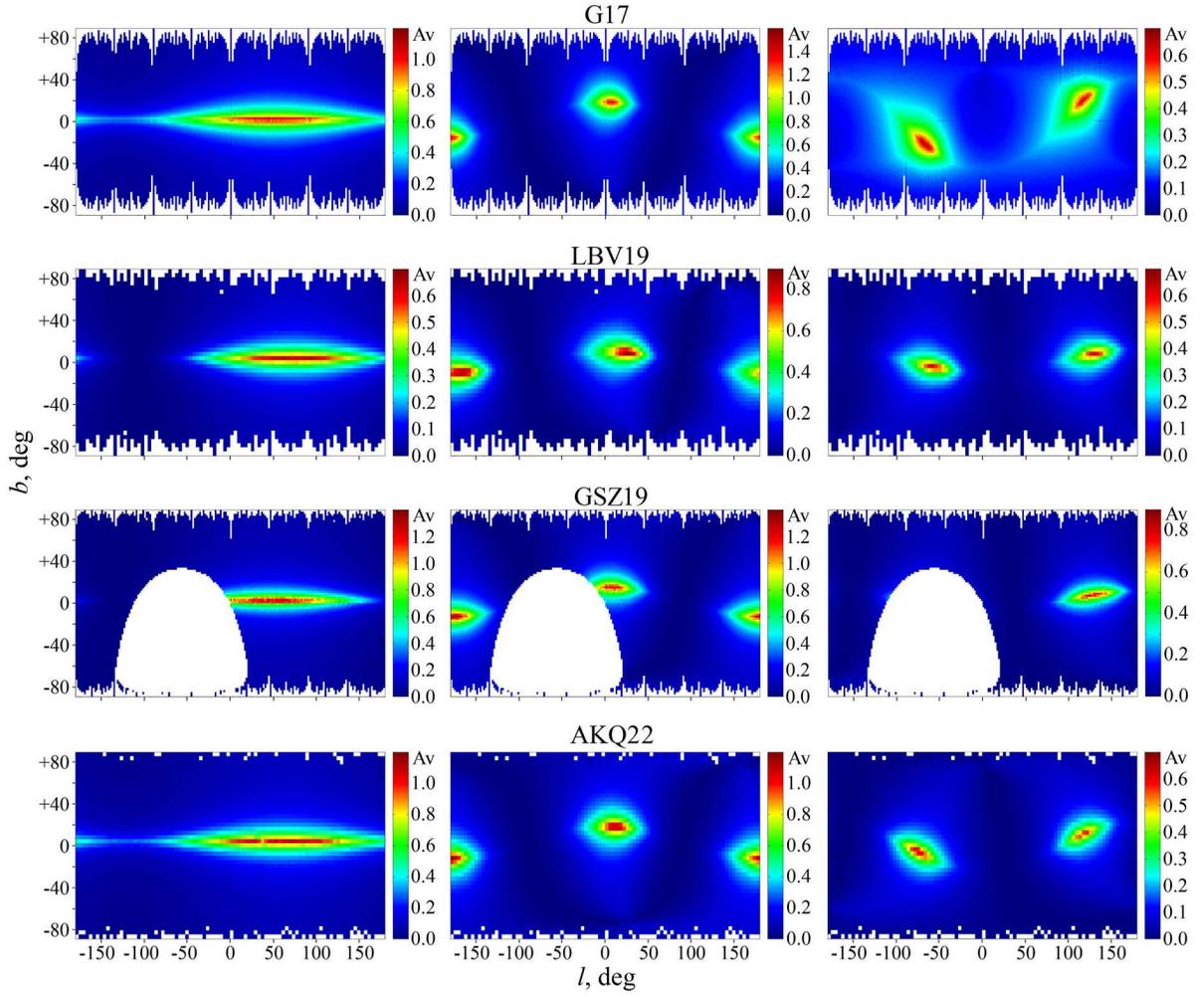}
\caption{Extinction $A_\mathrm{V}$ averaged in $2\times2$ deg$^2$ cells as a function of Galactic coordinates predicted by the model for the equatorial layer (the left column of graphs), 
the Gould Belt layer (the middle column of graphs), and the Cepheus-Chamaeleon layer (the right column of graphs) for different data sets. $A_\mathrm{V}$ is represented by the color 
scale on the right.
}
\label{egc}
\end{figure*}

\begin{figure*}
\includegraphics{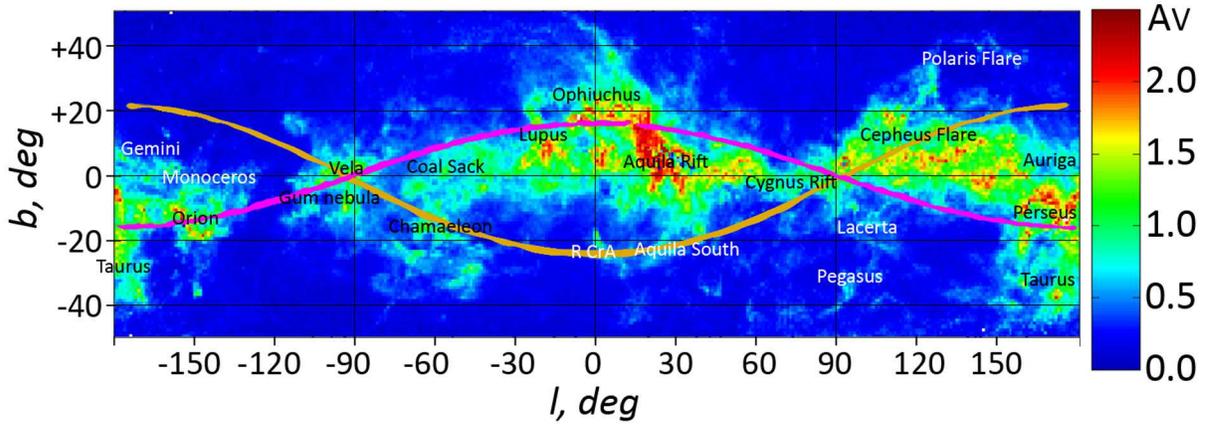}
\caption{Extinction $A_\mathrm{V}$ from AKQ22 averaged in $1\times1$ deg$^2$ cells as a function of Galactic coordinates. The projections of the midplanes of the Gould Belt and 
Cepheus-Chamaeleon layers are indicated by the purple and brown curves, respectively. The most noticeable cloud complexes within the nearest kiloparsec are inscribed. $A_\mathrm{V}$ 
is represented by the color scale on the right.
}
\label{clouds}
\end{figure*}

\begin{figure*}
\includegraphics{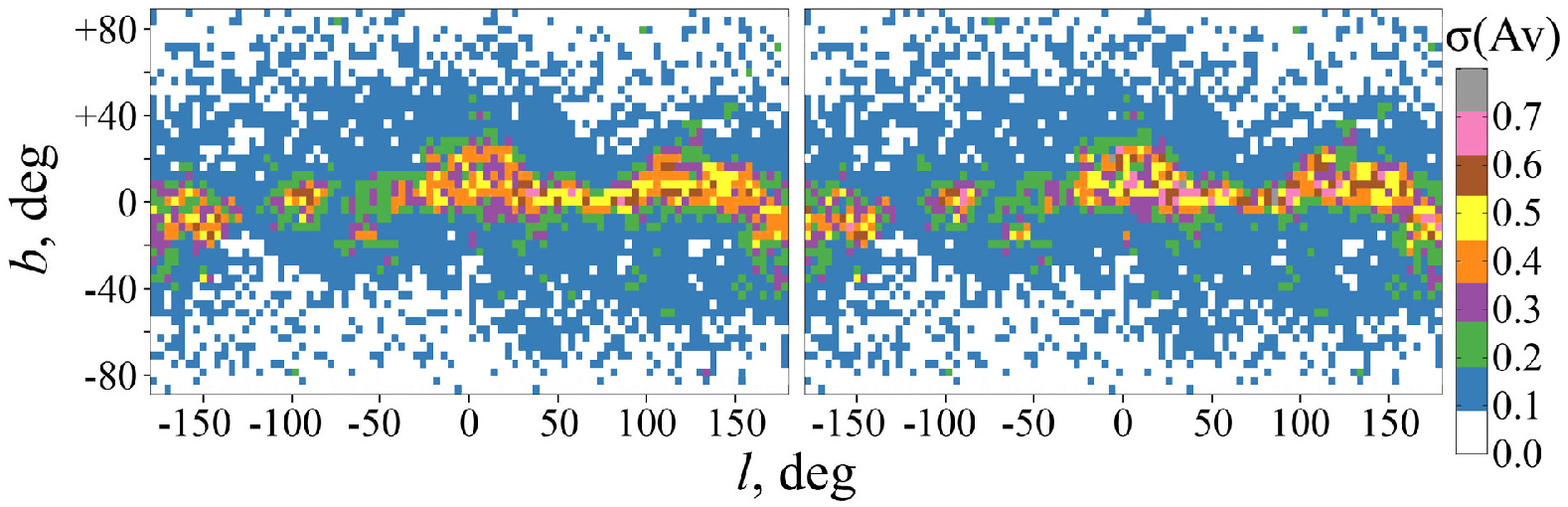}
\caption{Standard deviation of the $A_\mathrm{V}$ estimates from AKQ22 (left) and standard deviation of the `AKQ22 minus model' $A_\mathrm{V}$ residuals (right) calculated in $4\times4$
deg$^2$ cells as a function of Galactic coordinates. The values are represented by the color scale on the right.
}
\label{stdev}
\end{figure*}

Figure 1 shows the extinction averaged in $2\times2$ deg$^2$ cells as a function of Galactic coordinates: the $A_\mathrm{V}$ estimates from the data sets and the predictions
of our model are presented in the left and right columns, respectively. Since this comparison ignores the change in extinction with distance, it is incomplete , `two-dimensional' 
(a direct comparison of the observed and predicted extinctions is given below). Because of the approximately uniform distribution of the stars/points being used in space and the
increase in extinction with distance, Fig. 1 actually shows the extinction at a considerable distance, at a distance of several hundred parsecs from the Sun. However, this is enough 
to make sure that all of the data sets show qualitatively similar extinction variations over the sky that are successfully described by our model, though they are strongly smoothed
out. Only the disagreement between the data sets in the extinction estimates at high latitudes, where the stars/points with $R>500$ pc prevail, i.e., definitely outside the Galactic 
dust layer, is noticeable: from approximately zero extinction from the LBV19 data to maximum, essentially nonzero one from the G17 data. This disagreement between the data sets can
be explained and will be resolved in future studies (for a detailed discussion, see Gontcharov and Mosenkov 2018, 2021a, 2021b). A direct comparison of the data sets between themselves 
is beyond the scope of our study, except for the comparison of GSZ19 and AKQ22 in the Section `Discussion'.

Figure 1 can be compared with Fig. 7 from Gontcharov and Mosenkov (2021b). In particular, the predictions of the new version of our model (the right column in Fig. 1) can be compared 
with the predictions of the previous version of our model (the second row of graphs in Fig. 7 from Gontcharov and Mosenkov (2021b)). This comparison shows that the new version 
reproduces the observed extinction variations much more accurately. The Polaris Flare clouds near $l\approx125^{\circ}$, $b\approx+25^{\circ}$ and the southern part of the Chamaeleon 
cloud complex near $l\approx-60^{\circ}$, $b\approx-25^{\circ}$, which are clearly distinguished by all of the data sets, were reproduced by it particularly successfully.
This is already enough to justify the introduction of the Cepheus-Chamaeleon layer in our model.

Figure 2 shows the contribution of each layer to the total extinction estimates from our model as a function of Galactic coordinates. In other words, the right column in Fig. 1 is 
the sum of three columns in Fig. 2. It can be seen that the model shows a similar geometry of the layers when using different data sets, despite some differences in the solutions 
for different data sets in Table 1. The most important conclusion from Fig. 2 is that different data sets suggest very similar coordinates for the regions with a comparatively
high extinction at middle and high latitudes that are seen in the middle and right columns of graphs in Fig. 2.

Note that the very high eccentricities of the Gould Belt and Cepheus-Chamaeleon dust layers derived by us from all of the data sets suggest that a significant fraction of the data is 
contained in the small Aquila (Aquila Rift), Ophiuchus, Orion, Taurus, Perseus, Cepheus, and Chamaeleon cloud complexes. However, Fig. 2 shows that the remaining, equally 
significant fraction of the dust is distributed in vast space. Thus, the layers under consideration form extended containers of diffuse dust, while the known dust clouds are only 
the tip of the iceberg for these containers. Then, for any 3D analytical model of the extinction or the spatial dust distribution an accurate prediction of the extinction
at each point in space is much more important than the description of the known dust clouds. It should be emphasized that the semiminor and semimajor axes of the midplanes of the 
layers and the scale height of the layers do not point to the size of these dust containers, but are only parameters in the formal description of the dust medium. For example, in any
model with an exponential vertical distribution of dust its concentration is great even far from the midplane: a noticeable fraction, 5 per cent of the dust, is farther than
triple the scale height, i.e., in terms of Eq. (1) at $|Z|>3\,E_\mathrm{scale}$.

At middle and high latitudes the diffuse dust scattered in space that does not enter into large clouds is described by our model as belonging, to a large extent, to the Gould Belt 
and Cepheus-Chamaeleon layers. Consequently, these layers are important both as mathematical descriptions and as real dust containers when analyzing the extinction to extragalactic
objects at middle and high latitudes. For example, our averaged solution from Table 1 allocates 36 per cent, 57 per cent, and 7 per cent contributions to the total extinction toward the Galactic poles 
to the equatorial layer, the Gould Belt layer, and the Cepheus-Chamaeleon layer, respectively.

The dust density and the corresponding extinction in the Cepheus-Chamaeleon layer are probably lower than those in the remaining layers. This can be seen both from Fig. 2 and from 
our comparison of the constant terms in Table 1. This may be responsible for the great uncertainty and disagreement in the estimates of the layer tilt to the equator $C_\mathrm{tilt}$ based 
on different data sets. These uncertainties and disagreement are seen in Table 1 and Fig. 2. However, the fairly large values of the tilt $C_\mathrm{tilt}$ seem reasonable, since only such
large values can explain the existence of some clouds in the range $20^{\circ}<|b|<40^{\circ}$ that are seen in Fig. 1.

As noted previously, GSZ19 biases the estimates of some parameters due to the absence of data in the fourth and part of the third Galactic quadrants (Figs. 1 and 2). 
In particular, GSZ19 `does not see' the Chamaeleon cloud complex and, as a consequence, the estimates of the parameters for the Cepheus-Chamaeleon layer are biased. In addition,
this asymmetry in the spatial distribution of the GSZ19 data biases the estimate of the Gould Belt position with respect to the Sun.

Table 1 gives the positions of the centers of the Gould Belt and Cepheus-Chamaeleon layers with respect to the Sun. For the Cepheus-Chamaeleon layer this position has been found for 
the first time. The center of the Cepheus–Chamaeleon layer is offset toward the Galactic anticenter. The center of the Gould Belt layer 
($X=+5\pm5$, $Y=+75\pm13$, $Z=-3\pm19$, $R=76\pm24$ pc) turned out to be slightly closer to the Sun than follows from the estimates of Perryman (2009) and Bobylev (2014) based on
the spatial distribution of young stars (in the second quadrant, $R\approx150$ pc) and closer than follows from the previous version of our model (Gontcharov and Mosenkov 2021b): 
$X=-15\pm15$, $Y=+125\pm15$, $Z=-28\pm15$, $R=129\pm26$ pc. However, the spatial distribution of dust can differ from the distribution of young stars. Nevertheless, the main cause of
this discrepancy is the introduction of the Cepheus-Chamaeleon layer into the new version of our model. Indeed, the previous version, as most of the previous studies, treated all of 
the mid- and high-latitude clouds as parts of the Gould Belt. For example, Kirk et al. (2009) treated the Cepheus Flare clouds as a part of the Gould Belt, which is not confirmed 
by their positions at all.

For a better understanding of the geometry of the layers we reproduced the celestial extinction variations from AKQ22 in Fig. 3. Here the curves indicate the projections 
of the midplanes of the Gould Belt and Cepheus–Chamaeleon layers onto the celestial sphere obtained for AKQ22. We inscribed the large cloud complexes.

Figure 3 shows that the Chamaeleon Complex, the Corona Australis Complex around the star R~CrA, the Aquila South Complex, the Cepheus Flare, and the Polaris Flare, judging by their 
coordinates, belong to the Cepheus–Chamaeleon layer. Moreover, Fig. 3 shows that all of the large cloud complexes, except for the Pegasus cloud complex, are located along the midplanes 
of the three layers under consideration. This confirms that our model describes successfully the spatial distribution of the largest dust clouds within the nearest kiloparsec.

An interesting surprise is that the midplanes of the Gould Belt and Cepheus–Chamaeleon layers intersect the Galactic midplane approximately in one place, near $|l|\approx90^{\circ}$, 
i.e., near the $Y$ axis, although their tilts to the equator in Table 1 are slightly different: $B_\mathrm{tilt}=15^{\circ}\pm3^{\circ}$ and $C_\mathrm{tilt}=-28^{\circ}\pm13^{\circ}$. 
This forms a symmetric structure with two layers whose sides far from the equator are oriented identically, along the Galactic center--anticenter axis. Note that this symmetric 
structure was in no way assumed by our model from the outset.

Although the midplanes of the layers are approximately symmetric, the main clouds and the regions of enhanced extinction are not. This important difference between the Gould Belt 
and Cepheus-Chamaeleon layers is seen from Table 1 and Figs. 1-3. The maximum extinction in the Gould Belt is predicted by our model and is observed in the Belt regions farthest 
from the Galactic midplane. This is not the case for the Cepheus-Chamaeleon layer.

The standard deviations of the $A_\mathrm{V}$ residuals in Table 1 forG17 and LBV19 are much smaller than those for GSZ19 and AKQ22. To explain this, it is worth remembering that 
the residuals between the data and the model are caused by the natural fluctuations of the dust medium, the data errors, and the model errors (for a detailed analysis, see 
Gontcharov 2019). The fluctuations manifest themselves only in AKQ22 (in the data set with the extinctions for individual stars) and GSZ19 (on the high-resolution map). As noted
in the Introduction, in G17 and LBV19 the fluctuations are strongly smoothed out and do not contribute to the residuals.

Table 1 presents only the mean standard deviations of the $A_\mathrm{V}$ residuals over the entire sky. However, they vary dramatically with coordinates, particularly latitude. Figure 4 shows 
the standard deviation of the $A_\mathrm{V}$ estimates from AKQ22 and the standard deviation of the `AKQ22 minus model' $A_\mathrm{V}$ residuals as a function of Galactic coordinates. 
The first quantity includes the natural fluctuations of the dust medium and the data errors. Its typical values are $0.08^m$ and $0.37^m$ at high and low latitudes, respectively. The
errors at high latitudes declared by the authors of AKQ22 are comparable to the standard deviation of the extinction estimates. Therefore, to estimate the fluctuations of the medium, 
we invoked the result of direct observations -- the color dispersion for a sample of red clump giants. This sample was used to create the previous version of our model 
(Gontcharov and Mosenkov 2021b). The observed color dispersion includes the fluctuations of the dust medium and the dereddened color dispersion of clump giants due to the great 
variety of their own characteristics. The latter quantity is predicted fairly accurately by the models of the internal structure and evolution of stars, as shown by Gontcharov 
and Mosenkov (2021b). As a result, we obtained an estimate of the standard deviation $\sigma(A_\mathrm{V})=0.06$ at high latitudes due to the fluctuations of the dust medium.

The typical standard deviations of the `AKQ22 minus model' residuals are $0.09^m$ and $0.40^m$ at high and low latitudes, respectively, i.e., slightly larger than the typical 
standard deviations of the extinction estimates from AKQ22, since they also include the model errors. Having eliminated the contribution of the AKQ22 errors, we estimate the 
accuracy of the predictions of the extinction by our model for a star or a point in space, which changes from $0.07^m$ to $0.37^m$ at high and low latitudes, respectively.

The dominance of the fluctuations in both quantities presented in Fig. 4 also manifests itself in very similar variations of these quantities over the sky, particularly at high 
latitudes. A detailed comparison of these quantities in each cell allowed us to calculate their root-mean-square differences and their dependence on latitude. The values of 
$(0.09^2-0.08^2)^{0.5}=0.04^m$ and $(0.40^2-0.37^2)^{0.5}=0.15^m$ at high and low latitudes, respectively, are typical. These values characterize the accuracy of the model predictions
when ignoring the fluctuations of the medium. The fluctuations are ignored, for example, if the mean or median extinction for an extended object (a galaxy, a star cluster, a dust cloud, 
etc.) or for a small region of space is considered. In these cases, the predictions of our model must be very accurate. 

For comparison, note that with an uncertainty in the individual AKQ22 extinction of $\sigma(A_\mathrm{V})=0.15$ the extinction for an extended object at high latitudes
is determined with the same 0.04m accuracy as in our model only when averaging the extinction estimates from AKQ22 for $(0.15/0.04)^2=14$ stars that are projected onto this object. 
Consequently, only rare largest extragalactic objects are provided with the required number of AKQ22 stars being projected onto them to obtain a mean or median extinction estimate
as accurate as that using our model.

However, here we have in mind only the random accuracy. The systematic errors of the predictions of our model reproduce the systematic errors of the data sets used to calculate its 
parameters. In addition, our model, as any, has systematic errors due to its unavoidable incomplete reflection of the reality. The comparison of the predictions of our model with the
most accurate extinction estimates for star clusters and variable stars (see below) gives some idea of these systematic errors.

When analyzing Table 1, it is worth taking into account the fact that the systematic errors of the data and the model may not manifest themselves in the standard deviations of the 
residuals and the correlation coefficients being considered by us. Therefore, a smaller standard deviation or a larger correlation coefficient for some data set does not suggest 
that this data set is more accurate than the remaining ones.

\begin{figure*}
\includegraphics{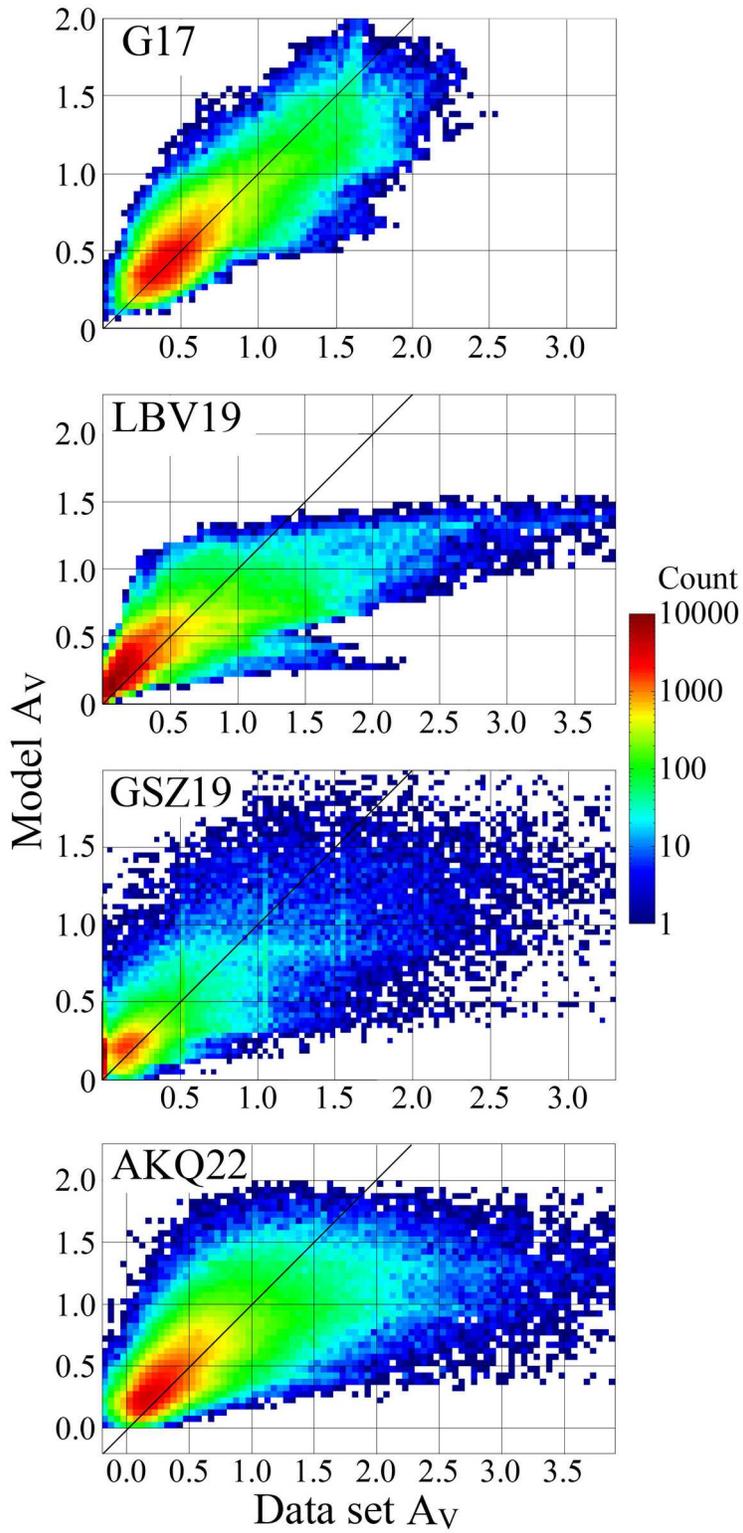}
\caption{Estimates of the extinction $A_\mathrm{V}$ from the data sets in comparison with the model predictions for randomly selected 256\,000 stars/points. The number of stars/points 
in each $A_\mathrm{V}$ cell is indicated by the color scale on the right.
}
\label{versus}
\end{figure*}

\begin{figure*}
\includegraphics{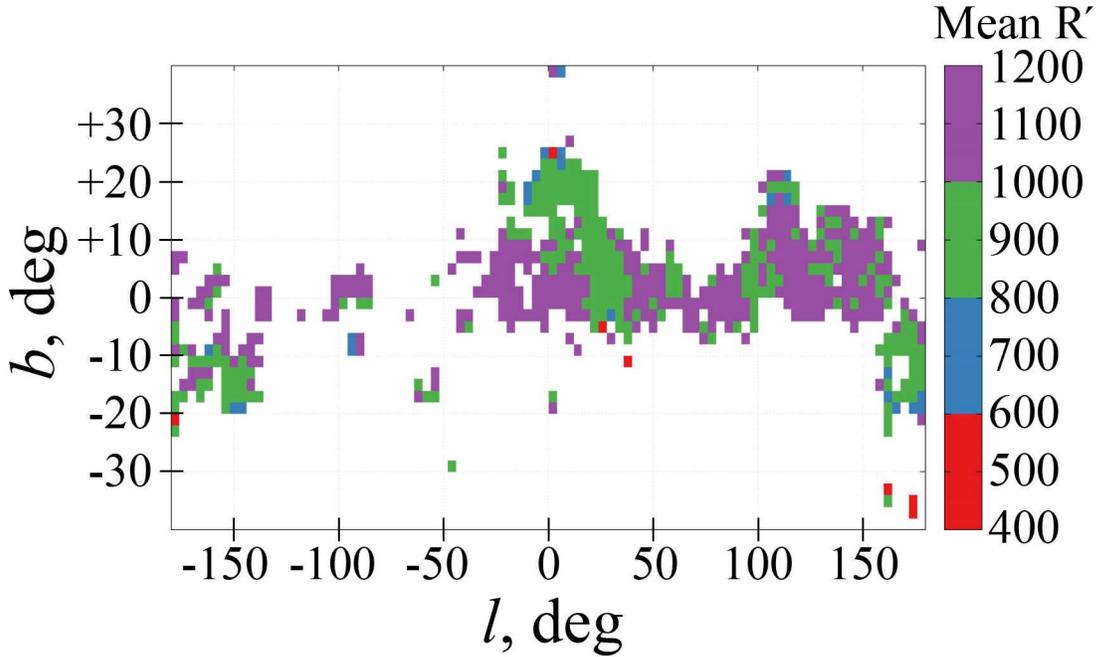}
\caption{Mean distance $R'\equiv(X^2+Y^2)^{0.5}$ (indicated by the color scale on the right) for 12\,921 stars from AKQ22 with $A_\mathrm{V}>2.3^m$ as a function of Galactic coordinates.
}
\label{outliers}
\end{figure*}

\begin{figure*}
\includegraphics{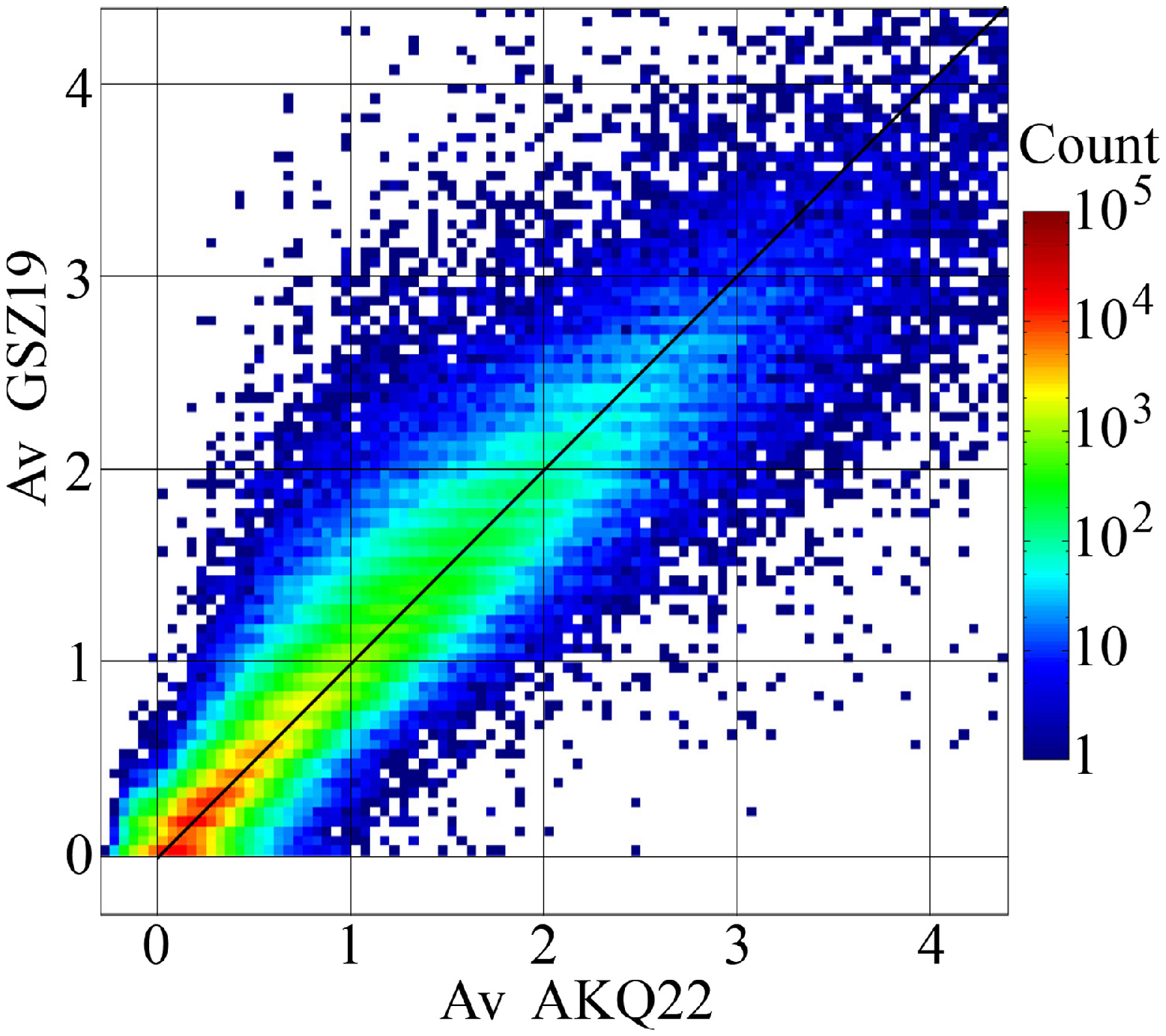}
\caption{Estimates of $A_\mathrm{V}$ from AKQ22 and GSZ19 for 664\,939 AKQ22 giants. The number of giants in each $A_\mathrm{V}$ cell is indicate by the color scale on the right.
}
\label{akq22gsz19}
\end{figure*}

\section*{DISCUSSION}

Let us compare the observed and predicted extinction estimates directly. In Fig. 5 we compared the randomly selected extinctions from the data sets with the corresponding predictions 
of our model. For all of the data sets most of the low extinctions are slightly overestimated by our model (the red cloud of cells is shifted upward relative to the bisector).
This is a result of the search for a balance between a large number of stars/points with a low extinction and a small number of stars/points with a very high extinction to obtain 
zero mean residual.

The stars/points with a very high extinction are predominantly at the edge of the space under consideration and apparently belong to Galactic structures outside the geometry of our 
model. An analysis of these stars/points allows the limitations of our model to be understood. Figure 6 shows the mean distance along the Galactic midplane defined by us as 
$R'\equiv(X^2+Y^2)^{0.5}$ for 12\,921 giants from AKQ22 with a high extinction $A_\mathrm{V}>2.3^m$ as a function of Galactic coordinates.\footnote{Only 12\,921 of the 993\,291 AKQ22 
giants (1.3 per cent) exhibit such a high extinction that is not predicted by our model.} It can be seen that such giants are
very rare up to $R'<800$ pc (the red and blue symbols). A comparison of Fig. 6 with Fig. 3 shows that these giants belong to the Gould Belt clouds and the Aquila South 
cloud.\footnote{Figures 3 and 6 have different scale heights along the vertical axis.} 
The giants with $A_\mathrm{V}>2.3^m$ are more numerous at $800<R'<1000$ pc (the green symbols in Fig. 6), but they still belong predominantly to the equatorial layer and the Gould Belt. 
Most of the giants with $A_\mathrm{V}>2.3^m$ have $R'>1000$ pc (violet symbols) and, in contrast to the closer giants, belong to the Cepheus Flare, Cygnus Rift, Vela Molecular
Ridge, Monoceros Complex clouds and the clouds toward the Galactic center, as can be seen from our comparison of Fig. 6 with Fig. 3.

Note in this connection that all of the data sets give consistent estimates of the semimajor axis for the Gould Belt midplane with the mean $B_\mathrm{major}=743\pm61$ pc. 
This suggests that the Belt dust is exclusively within the nearest kiloparsec, although it also obscures more distant objects. In contrast to the Gould Belt dust, only part of 
the dust in the Cepheus-Chamaeleon layer is within the space being considered by us, as follows from the consistent estimates of the semimajor axis $C_\mathrm{major}>1000$ pc for
the data sets. This is consistent with other estimates of the distance to the Cepheus Flare cloud complex (see Kun et al. 2008; Chen et al. 2020, and references therein).

Returning to the stars/points with a low extinction in Fig. 5, note their dichotomy for GSZ19. Just as for the remaining data sets, the set of points (the orange clump on the graph) 
shows a normal or lognormal distribution in both observed and predicted extinctions. However, another set of points (the red vertical column on the graph) shows zero observed and non-zero 
predicted extinctions. It is important that this extinction does not refer to the space near the Sun, since GSZ19 gives no extinctions within $\approx300$ pc from the Sun, as noted in 
the Section `Data'. The points with zero extinction from GSZ19 are at high latitudes far from the Galactic midplane. Thus, at these points GSZ19 erroneously shows zero extinction across 
the whole Galactic dust layer. This dichotomy, i.e., both zero and significantly nonzero extinction estimates at high latitudes, was revealed by Green et al. (2018), who noted that
for line of sights with a small volume of data the first version of their map (Green et al. 2015) has a tendency to assign zero reddenings in those cases where the true reddening is 
less that a few hundredths of a magnitude. It can be seen that this feature is retained in the latest version of the GSZ19 map and manifests itself when estimating the extinction
for high-latitude clouds. For example, Gontcharov and Mosenkov (2021a) found that GSZ19 obtained reliable, significantly nonzero extinction estimates \emph{within} high-latitude Galactic 
globular clusters using photometry for dozens of cluster member stars along each line of sight. However, for the sky regions \emph{around} these clusters GSZ19 obtained unreliable
zero or almost zero extinction estimates due to the use of only a few isolated stars along each line of sight.

It is important that these erroneous zero extinctions distort severely some of the parameters of our model found from GSZ19. For example, a large fraction of zero extinctions far 
from the Galactic midplane leads to an underestimation of the mean extinction there. This leads to an unrealistically rapid decrease in extinction with $|Z|$ and, consequently, 
to an underestimation of $E_\mathrm{scale}$ from GSZ19 (note that the underestimation of $E_\mathrm{scale}$ from LBV19 is caused by the constraint |$|Z|<400$ pc). Moreover, to follow an exponential decrease 
in extinction with $|Z|$, our model compensates for the extinction underestimation far from the Galactic midplane by the extinction overestimation near this midplane, and this leads 
to an overestimation of $E$ and $E_\mathrm{amplitude}$ from GSZ19. This remark shows the relationship between the parameters of our model that characterize the dust density and emphasizes once 
again the previously mentioned division of the model parameters into two categories. We see that the derived dust density characteristics can be plagued by significant systematic 
errors and require invoking more accurate data sets.

Having assumed that in the regions of space with a low extinction the AKQ22 estimates are systematically more accurate than the GSZ19 estimates, we can replace the zero GSZ19 
extinctions with more realistic estimates. In Fig. 7 we compare the AKQ22 and GSZ19 extinctions for 664\,939 giants having extinction estimates from AKQ22. It can be seen
that the minimum GSZ19 extinctions are systematically underestimated compared to AKQ22 by $\Delta A_\mathrm{V}=0.12^m$. This correction would remove the mentioned dichotomy and 
would combine all of the low extinction estimates from GSZ19 around the median $A_\mathrm{V}=0.13^m$.

At $|b|>62^{\circ}$ all of the data sets show invariable statistical characteristics. The extinction in this range of high latitudes far from the Galactic midplane may be treated as 
the extinction across the whole Galactic dust half-layer above or below the Sun. LBV19, GSZ19, AKQ22, and G17 estimate this extinction as $A_\mathrm{V}=0.07^m$, $0.11^m$, $0.12^m$ and
$0.20$, respectively (given the corrected zero GSZ19 extinctions). A summary of other estimates of this extinction and an accompanying analysis of the systematic errors were given by 
Gontcharov and Mosenkov (2021a).

The most popular 2D reddening maps from the dust emission observations by Schlegel et al. (1998) and Meisner and Finkbeiner (2015) estimate the extinction across the whole Galactic 
dust half-layer above or below the Sun as $A_\mathrm{V}=0.06^m$ and $0.05^m$, respectively, if the CCM89 extinction law with $R_\mathrm{V}=3.1$ is adopted, and, consequently, 
underestimate the extinction at high latitudes far from the Galactic midplane compared to all of the data sets under consideration. The estimates by Schlegel et al. (1998) and
Meisner and Finkbeiner (2015) are based on the dust emission calibrations from the reddening $E(B-V)$ for a set of stars and elliptical galaxies. As noted by the authors themselves, 
the accuracy of these calibrations is low: for the map of Schlegel et al. (1998) it actually leads to an estimate of $A_\mathrm{V}=0.06\pm0.09^m$ for the extinction across the whole 
Galactic dust half-layer above or below the Sun. Moreover, by calibrating the dust emission from galaxy counts, Schlegel et al. (1998) obtained an estimate ($A_\mathrm{V}=0.12\pm0.09^m$) 
that is twice as large as that from the galaxy color calibration. It can be seen that the first and, a fortiori, the second estimates are formally consistent with the mentioned 
estimates from LBV19, GSZ19, AKQ22, and G17 if their uncertainties are taken into account.

Apparently, of all the estimates of the extinction across the Galactic dust layer, the AKQ22 estimate is most accurate, as the result of a recent study using the smallest number 
of assumptions and calibrations based on the most accurate parallax estimates and photometry for a record large number (hundreds of millions) of stars. The AKQ22 estimate forces us
not only to revise the extinction estimates for high latitude extragalactic objects, but also requires to explain how a substantial amount of dust ended up far from the Galactic 
midplane. Our model is just a step in this explanation, assuming that the dust far from the Galactic plane is contained predominantly in dust structures tilted to this plane. 
Thus, our model reduces the question about the appearance of dust far from the Galactic midplane to the question about the formation and evolution of the Gould Belt and the
Cepheus–Chamaeleon dust layer.

However, the new estimates of the extinction across the Galactic dust layer apparently do not require a radical revision of the characteristics of extragalactic objects associated 
with them. For example, the difference between the Schlegel et al. (1998) and AKQ22 estimates can be removed to a first approximation using a constant correction over the entire
sky, $\Delta E(B-V)=0.02$. Such a small correction, for example, will hardly change the dispersion, the median, the mean, and other color characteristics of the elliptical galaxies 
without current star formation that form a compact `red sequence' with a comparatively small color dispersion on the color–absolute magnitude diagram (Chilingarian and Zolotukhin 2012;
Chilingarian et al. 2017). Indeed, the red-sequence color distortion by $\Delta B-V=0.02$ due to the reddening underestimation is apparently smaller than the uncertainty in the 
predictions of the present-day models. In addition, our estimate of the extinction fluctuations at high latitudes $\sigma(A_\mathrm{V})=0.06$ makes a minor contribution, 
$\sigma(E(B-V))=0.02$, to the red sequence color dispersion. Moreover, for galaxies, as extended objects, this dispersion is significantly smoothed out.

\begin{table}
 \centering
\def\baselinestretch{1}\normalsize\normalsize
\caption[]{Bayesian information criterion for four data sets (random subsamples of 256\,000 stars/points each) and three versions of our model (the new version with two layers, with 
three layers, and the fourth version of Gontcharov and Mosenkov (2021b))
}
\label{bic}
\begin{tabular}[c]{lcccc}
\hline
\noalign{\smallskip}
Model version       & G17 & LBV19 & GSZ19 & AKQ22 \\
\hline
\noalign{\smallskip}
Fourth version    & $-860\,382$ & $-472\,487$ & $-464\,302$ & $-451\,615$  \\ 
Two layers        & $-835\,303$ & $-695\,350$ & $-554\,846$ & $-509\,382$ \\ 
Three layers      & $-890\,251$ & $-733\,623$ & $-568\,441$ & $-524\,203$ \\ 
\hline
\end{tabular}
\end{table}


To check whether it is necessary to use three dust layers in our model, we used the same data sets to calculate the model parameters, but without the Cepheus-Chamaeleon layer or 
only with one equatorial layer. In addition, we made sure that the new version of our model has an advantage over the previous version presented by Gontcharov and Mosenkov (2021b). 
In all these cases, we obtained standard deviations of the AV residuals higher by a few hundredths, while the correlation coefficients are lower than those presented in Table 1 for 
the new version of our model with three layers by a few hundredths.

Furthermore, we additionally checked the advantages of the new version of our model by taking into account different numbers of parameters in different versions of our model. We used 
the Bayesian information criterion (BIC). BIC calculates the model efficiency with a penalty for the number of parameters. Using BIC, it can be seen whether the introduction
of additional parameters is justified. For data $x_i$ and model predictions $\hat{x}_i$ BIC is calculated as
$$
\mathrm{BIC} = n \ln \left(\frac{1}{n}\sum_{i=1}^n(x_i-\hat{x}_i)^2\right) + k \ln(n),
$$
where $k$ is the number of parameters and $n$ is the number of stars/points under consideration. A smaller BIC means a better correspondence between the model and the data. 
The derived values of BIC are presented in Table 2. In all cases, the new version of our model with three layers and 29 parameters gives smaller BIC (all of the values in Table 2 are
negative!), i.e., it looks better than the version with two layers (without the Cepheus-Chamaeleon layer) and 17 parameters and the previous version of our model with 18 parameters. 
Thus, the inclusion of the separate Cepheus-Chamaeleon layer is justified.

We approximated the four data sets being considered by us by the 3D models of Arenou et al. (1992), Drimmel et al. (2003), Am\^ores and L\'epine (2007), and the simplest model mentioned 
in the Introduction. All these models give standard deviations of the $A_\mathrm{V}$ residuals higher by a few hundredths, while the correlation coefficients are lower than those in our 
new model by a few hundredths. This is expectable after our analysis of the models when they were applied by Gontcharov and Mosenkov (2021b) to a large sample of Gaia giants.

The factors  $R$, $\min(R,B_\mathrm{Rmax})$ and $\min(R,C_\mathrm{Rmax})$ in Eqs. (7)--(9) suggest a proportionality of the extinction and the distance near the Sun, i.e., a constancy
of the dust density. Our results show that this suggestion is plausible. This contradicts the popular view of the Gould Belt as a gas-dust torus with a region of reduced gas and dust 
density at its center. This region has a radius $\approx100$ pc around the Sun and is called the Local Bubble. Until recently, the low accuracy of reddening/extinction and distance
measurements has not allowed one to answer the question of whether the Local Bubble is indeed a region of reduced density of the medium. Gontcharov and Mosenkov (2019) showed that 
interstellar polarization measurements in combination with Na{\sc~I} and Ca{\sc~II} absorption line equivalent width measurements delineate more reliably the Bubble boundaries than 
do reddening or extinction measurements and, at the same time, show that the Bubble is not a region of reduced density of the medium, but a region of its enhanced ionization. 
An analysis of the whole variety of measurements inside and outside the Bubble (reddening, extinction, polarization, and absorption lines) by Gontcharov and Mosenkov (2019) showed
that only the LBV19 map displays the Bubble as a region of reduced dust density, and this is most likely a manifestation of the LBV19 systematic errors. This study allows us to check 
which data sets show a drop in the dust density in the Bubble.

To check this, we calculated the parameters of our model using G17, LBV19, and AKQ22 (GSZ19 gives no extinctions near the Sun), but nulled all of the extinctions in the Local Bubble, 
i.e., admitted a region of reduced dust density near the Sun.\footnote{Nulling the extinctions within 40 pc of the Sun does not change the standard deviations of the $A_\mathrm{V}$ 
residuals and the correlation coefficients at all, since the extinctions here anyway are very close to zero.} We varied the Bubble radius from 0 to 200 pc. It turned out that only
for LBV19 and only within 80 pc of the Sun nulling the extinctions reduced the standard deviation of the AV residuals and increased the correlation coefficient, i.e., improved the 
correspondence between the data and the model. Thus, LBV19, in contrast to G17 and AKQ22, suggests a region of reduced dust density within 80 pc of the Sun.

\begin{figure*}
\includegraphics{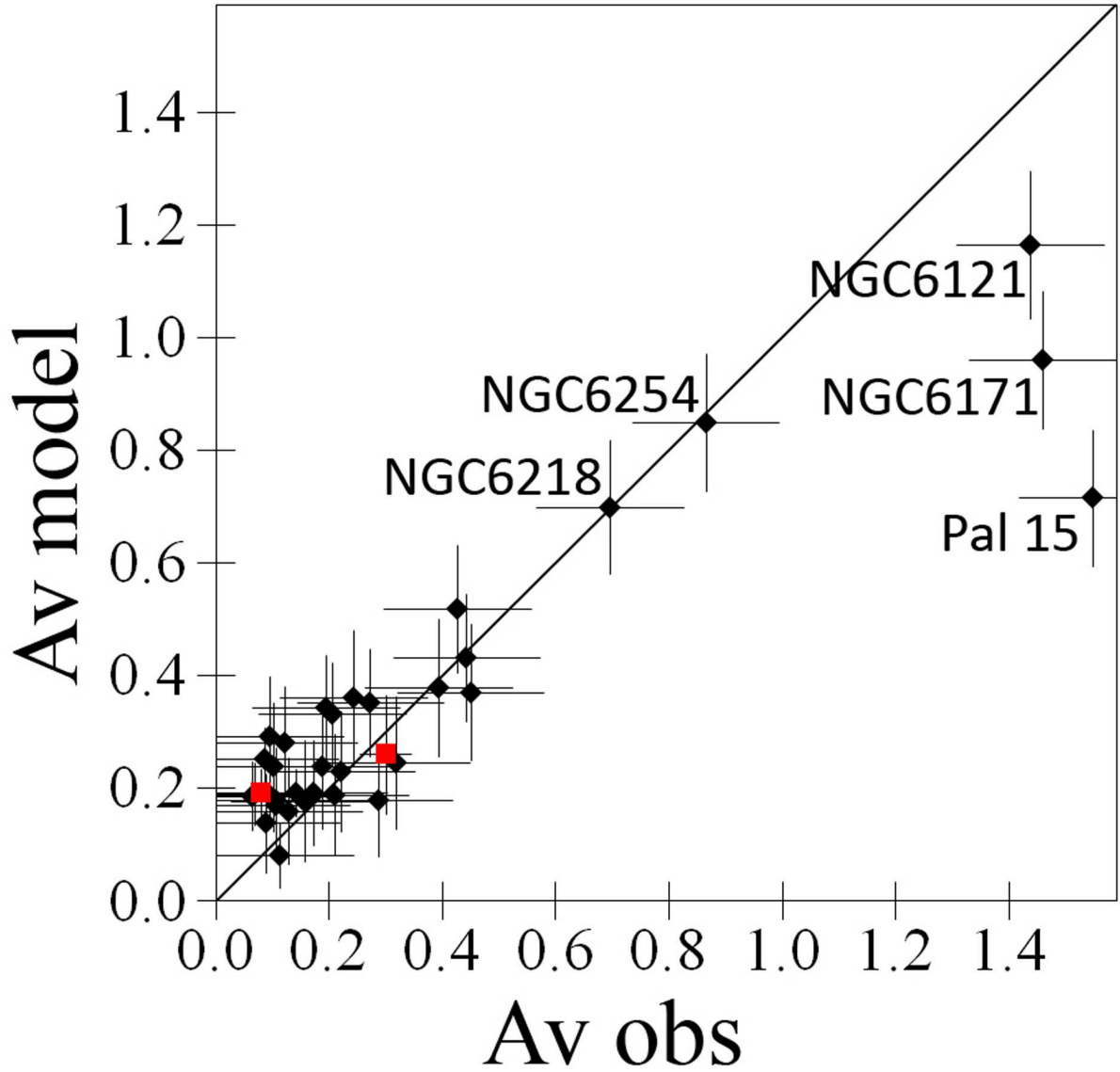}
\caption{Estimates of $A_\mathrm{V}$ for Galactic globular clusters from the literature in comparison with the predictions of our model (black diamonds). The estimates by 
Clementini et al. (2022) are marked by the red squares.
}
\label{gc}
\end{figure*}

\begin{figure*}
\includegraphics{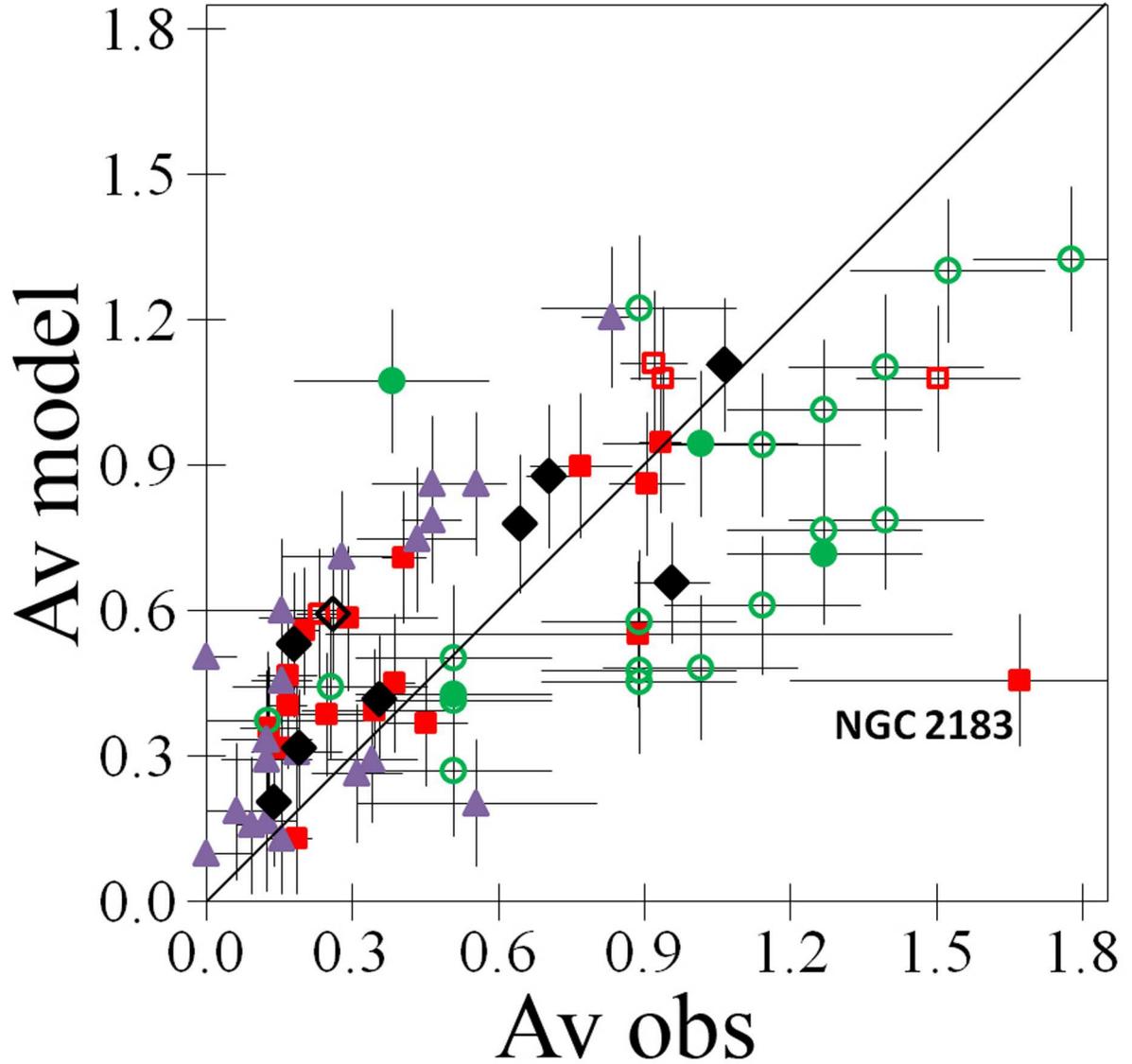}
\caption{Estimates of  $A_\mathrm{V}$ for open clusters from the literature in comparison with the predictions of our model: from Niu et al. (2020) - black diamonds, 
Monteiro et al. (2020) - red squares, He et al. (2021) - green circles, Jackson et al. (2022) - violet triangles. The filled and open symbols mark the clusters within 
$(X^2+Y^2)^{0.5}<1000$ and $1000<(X^2+Y^2)^{0.5}<1200$ pc, respectively.
}
\label{openclusters}
\end{figure*}

\begin{figure*}
\includegraphics{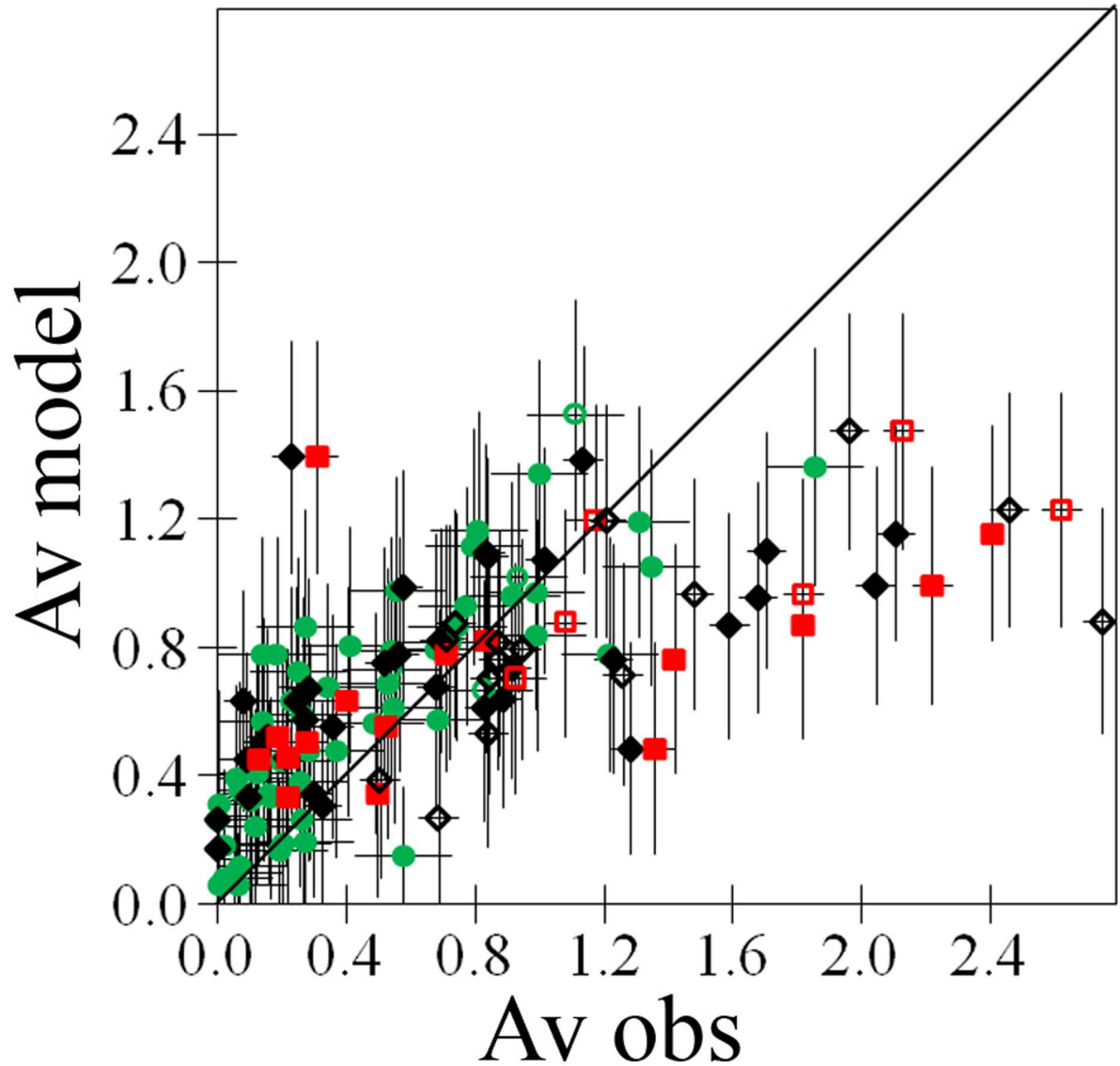}
\caption{Estimates of $A_\mathrm{V}$ for Cepheids and supergiants from the literature in comparison with the predictions of our model:
Cepheids from Kovtyukh et al. (2008) - black diamonds, Cepheids from Lazovik and Rastorguev (2020) - red squares, supergiants from Kovtyukh et al. (2008) - green circles. 
The filled and open symbols mark the stars within 
$(X^2+Y^2)^{0.5}<1000$ and $1000<(X^2+Y^2)^{0.5}<1200$ pc, respectively.
}
\label{cepheids}
\end{figure*}

\begin{figure*}
\includegraphics{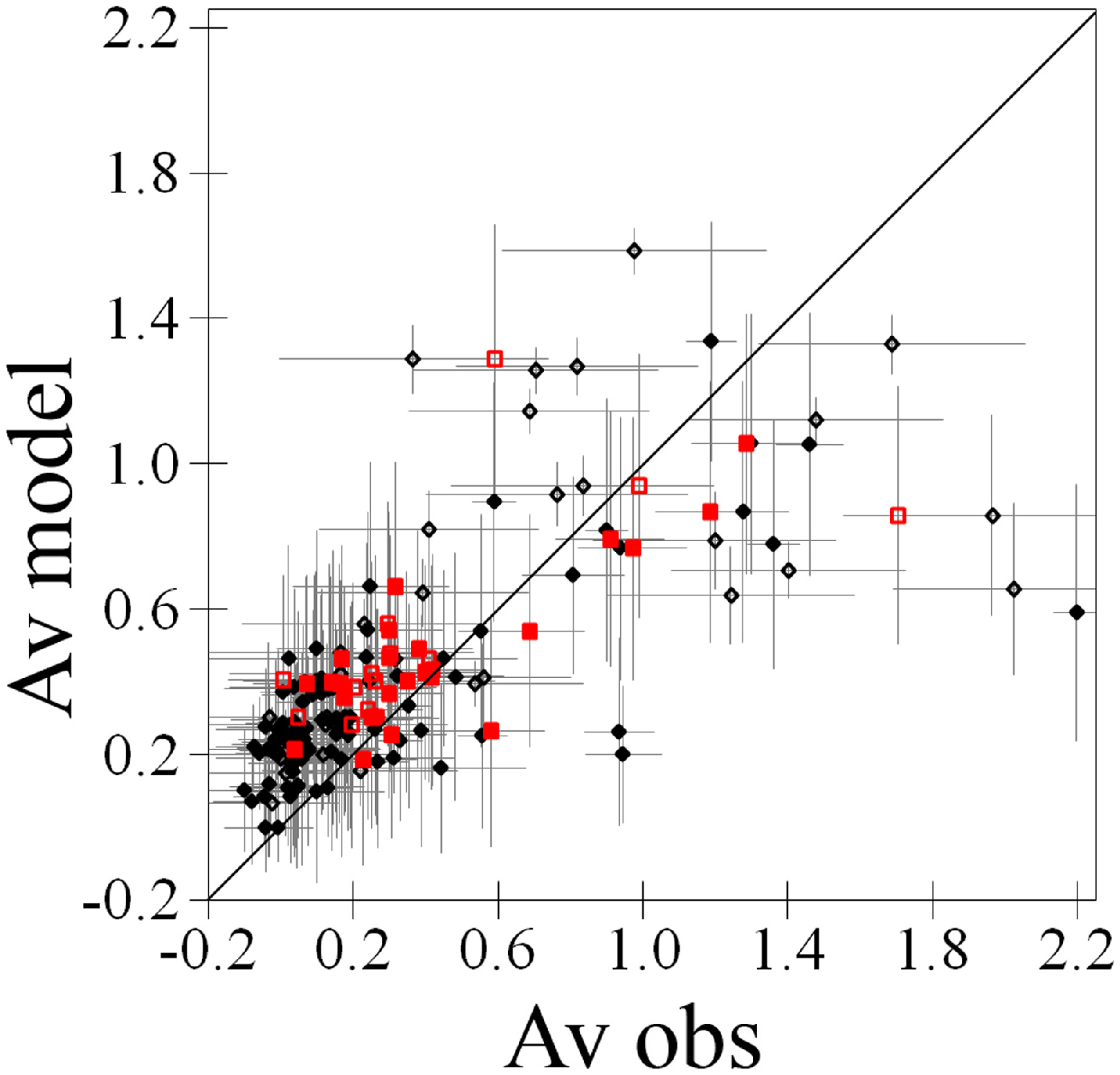}
\caption{Estimates of $A_\mathrm{V}$ for RR~Lyrae variable stars from the studies by Dambis et al. (2013, red squares) and Clementini et al. (2022, black diamonds) in comparison 
with the predictions of our model. The filled and open symbols mark the stars within $(X^2+Y^2)^{0.5}<1000$ and $1000<(X^2+Y^2)^{0.5}<1200$ pc, respectively.
}
\label{rrlyrae}
\end{figure*}

\section*{APPLICATION TO CLUSTERS AND VARIABLE STARS}

To estimate the quality of our model, we compared its predictions with the most accurate extinction estimates for open and globular clusters as well as for RR Lyrae variable stars 
and classical Cepheids. We selected the estimates obtained for variable stars based on the period--luminosity--metallicity relations and for clusters by comparing the color--magnitude
diagrams with theoretical isochrones. In addition, we considered the extinction estimates for a set of RR Lyrae variables within the nearest kiloparsec, in the globular cluster NGC\,288, 
and in the dwarf galaxy Ursa Major II obtained by Clementini et al. (2022) from their analysis of the Gaia DR3 data based on the period--color--amplitude relation for fundamental mode
variable stars. All of the reddening or extinction estimates used by us were converted to the $A_\mathrm{V}$ estimates using the CCM89 extinction law with $R_\mathrm{V}=3.1$. 
Our comparison of the estimates for the same objects from different publications revealed differences between them that occasionally exceeded considerably the uncertainties declared 
by the authors. This suggests that the authors underestimated the systematic errors. Note that the possible spatial variations of the extinction law disregarded by us can contribute 
to the disagreement between the estimates.

For the model predictions we used the set of parameters from Table 1 for AKQ22 as the one based on the most extensive, accurate, and new material. The accuracy of the predictions 
depends on the latitude. As noted previously, our model gives inaccurate predictions for both point and extended objects far from the Sun at low latitudes, at 
$(X^2+Y^2)^{0.5}>1200$ pc and $|b|<24^{\circ}$. Our comparison with the estimates for clusters and variable stars confirmed this. Below such clusters and stars are not considered, 
except for some illustrative examples.

Our comparison for 34 Galactic globular clusters and the dwarf galaxy Ursa Major II is presented in Fig. 8. We considered all of the known globular clusters with $|b|>24^{\circ}$ 
and within 40 kpc of the Sun. For more distant clusters the photometry on the color–magnitude diagram is not deep enough for accurate estimates. Apart from them, we considered
three clusters to show the limitations of our model: NGC\,6121 and NGC\,6171 are close to the Galactic equator ($b\approx16^{\circ}$ and $b\approx23^{\circ}$, respectively)
and, at the same time, are far from the Sun (2.2 and 6.4 kpc, respectively), while Palomar 15 is too far from the Sun ($\approx45$ kpc). We used the reddening/extinction estimates 
from the studies by Wagner-Kaiser et al. (2016, 2017), Dotter et al. (2010, 2011), Bellazzini et al. (2002), Koch and McWilliam (2014), Hamrick et al. (2021), Ortolani and Gratton (1990),
and Hamren et al.(2013), giving preference to them precisely in this order if there are estimates for a cluster in different publications.

For all of the reddening/extinction estimates for globular clusters taken from the literature we adopted the most realistic, in our view, estimate of the uncertainty 
$\sigma(A_\mathrm{V})=0.13$. For the estimates from Clementini et al. (2022) we adopted the uncertainties declared by the authors. Note that although there are reddening/extinction 
estimates for all of the known Galactic globular clusters in the most popular database of globular clusters by Harris (1996, the latest version 2010), many of them either were
calculated before 1990 using isochrones with a very low accuracy or were taken from the Schlegel et al. (1998) map that underestimates the extinction at middle and high latitudes, 
as note previously.

Figure 8 shows good agreement between the predictions of our model and the estimates from the literature for most of the globular clusters. However, the isolated group of clusters 
demonstrates some disagreement between the observed, $A_\mathrm{V}\approx0.10$, and predicted, $A_\mathrm{V}\approx0.25$, extinctions. This suggests a possible overestimation of 
the low extinctions by our model.

Despite the large number of recent publications with extinction estimates for open clusters, the difficulties in revealing cluster members, the small number of members, the extinction 
and metallicity degeneracy, the possible abundance of dust inside the cluster, and other difficulties reduce the accuracy of the estimates. Examples of comparatively accurate
estimates for clusters in completely different directions were given by Niu et al. (2020), Monteiro et al. (2020), He et al. (2021), and Jackson et al. (2022). Figure 9 shows these 
extinction estimates in comparison with the predictions of our model. For the estimates of He et al. (2021) we adopted the uncertainty $\sigma(A_\mathrm{V})=0.2$. 
We see acceptable agreement both for the clusters with $(X^2+Y^2)^{0.5}<1000$ pc and for many of the clusters with $1000<(X^2+Y^2)^{0.5}<1200$ pc. Just as for globular clusters, 
the disagreement between the observed and predicted estimates for low extinctions argues for some overestimation of them by our model.

Many of the clusters for which the observed extinction is much higher than the predicted one contain star-forming regions and, consequently, dust inside the cluster. This dust is 
ignored by our model. Therefore, these clusters known to us were excluded from consideration. As an example, we left the cluster NGC\,2183 embedded in a gas–dust cloud, which is
marked in Fig. 9. 

Figure 10 shows the extinction estimates for Cepheids and supergiants from Kovtyukh et al. (2008) and Cepheids from Lazovik and Rastorguev (2020) in comparison with 
the predictions of our model. For supergiants we adopted the uncertainty $\sigma(A_\mathrm{V})=0.15$. The disagreement between the extinction estimates for common Cepheids from these
two papers often exceeds considerably the declared uncertainties. Just as for clusters, it can be seen that the low extinctions may have been overestimated by our model. 
For an observed $A_\mathrm{V}>1.2$ the pattern of disagreement between the observations and the predictions changes for Cepheids, but, apparently, not for supergiants. 
Consequently, this disagreement cannot be completely explained by the errors of our model.

Figure 11 shows the extinction estimates for 39 and 131 RR Lyrae variables from the studies by Dambis et al. (2013) and Clementini et al. (2022), respectively, in comparison with 
the predictions of our model. We did not use the Dambis et al. (2013) estimates for stars with $|b|>25^{\circ}$, since they reproduce the estimates from the Drimmel et al. (2003)
model. For the Dambis et al. (2013) estimates we adopted the uncertainty $\sigma(A_\mathrm{V})=0.15$. The Clementini et al. (2022) data were cleaned of the extraneous stars 
erroneously identified in the Gaia project as RR Lyrae variables using the absolute magnitude constraints for the broad Gaia band $0<M_\mathrm{G}<1.43$. Note that for many of the 
RR Lyrae variables Clementini et al. (2022) obtained negative extinctions. Even the median of the extinction for RR Lyrae stars at $|b|>50^{\circ}$ is negative: $A_\mathrm{V}=-0.01$.
Moreover, according to Clementini et al. (2022), at high latitudes the extinction decreases with distance. We selected nearby RR Lyrae stars as stars with a Gaia parallax larger than 
0.5 mas and a relative parallax accuracy better than 0.1. For them the median of the extinction is also very small: $A_\mathrm{V}=0.02$. Thus, the Clementini et al. (2022) estimates 
actually suggest zero extinction through the whole dust layer across the Galaxy near the Sun. Accordingly, this unrealistic estimate creates the discrepancy between the observations 
and the predictions of our model visible in Fig. 11. Apparently, the calibrations for RR Lyrae variables should be revised, while the general question about the extinction through the
whole dust layer across the Galaxy requires special attention.

\section*{CONCLUSIONS}

In this study we presented a new version of our 3D analytical model of spatial interstellar extinction variations. This is the fifth version of the model after those presented by 
Gontcharov (2009), Gontcharov (2012b), Gontcharov and Mosenkov (2019), and Gontcharov and Mosenkov (2021b).
This version describes the 3D dust distribution within the nearest kiloparsec in three overlapping layers: along the Galactic midplane, in the Gould Belt, and in the layer passing 
through the Cepheus and Chamaeleon cloud complexes. Each layer has exponential vertical and sinusoidal longitudinal dust density variations. The equatorial layer is infinite along 
the $X$ and $Y$ axes and is offset with respect to the Sun along the $X$ axis. The remaining layers have finite elliptical midplanes offset with respect to the Sun along all three 
coordinate axes.

The midplanes of the layers turned out to pass near all of the largest cloud complexes within the nearest kiloparsec: Gemini, Monoceros, Vela, Coal Sack, Cygnus Rift, and Auriga 
for the equatorial layer, Taurus, Orion, Gum Nebula, Lupus, Ophiuchus, Aquila Rift, Lacerta, and Perseus for the Gould Belt, and Chamaeleon, Corona Australis, Aquila South,
Cepheus Flare, and Polaris Flare for the Cepheus-Chamaeleon layer. However, what is more important, we obtained consistent estimates of 29 parameters of our model when using the 
four most accurate and extensive data sets for their calculation: the extinctions for 993\,291 giants from AKQ22 obtained from the Gaia results and the three G17, LBV19, and GSZ19
3D reddening maps. This justifies the possibility of using our model to predict the extinction at any point of the diffuse medium within the nearest kiloparsec and far beyond it at 
high latitudes ($|b|>24^{\circ}$). The accuracy of predicting the extinction $A_\mathrm{V}$ by our model for a star or a point in space ranges from $0.07^m$ to $0.37^m$ for high and 
low latitudes, respectively. This accuracy is limited by the natural fluctuations of the interstellar dust medium. Our model gives the mean or median extinction for extended objects 
or small regions of space, when the fluctuations of the medium are ignored, with a very high accuracy from $0.04^m$ to $0.15^m$ for high and low latitudes, respectively.

According to our model, the Gould Belt and Cepheus-Chamaeleon dust layers tilted to the Galactic midplane make a major contribution to the extinction at high latitudes. 
Thus, our model describes quantitatively the properties of the dust containers that can provide a comparatively high extinction, $A_\mathrm{V}=0.12^m$, through the whole Galactic 
dust half-layer above or below the Sun obtained recently in AKQ22 from Gaia results.

\section*{ACKNOWLEDGMENTS}

This study was supported by the Russian Science Foundation (project no. 20-72-10052).

We thank M.~Khovrichev for his help in accessing the Internet resources and A.~Dryanichkin, E.~Evseev, M.~Garanina, E.~Gordeev, P.~Popov, and N.~Svetlova
for their help in the calculations.

We are grateful to the referees for their useful remarks. 

In this study we used resources from the Strasbourg Astronomical Data Center \\
(http://cdsweb.ustrasbg.fr), including the SIMBAD database and the X-Match service. 
This study uses the Filtergraph online data visualization system (Burger et al. 2013, https://filtergraph.com). 
This study uses data from the Gaia mission of the European Space Agency (https://www.cosmos.esa.int/gaia) processed by the Data Processing and Analysis Consortium \\
(DPAC, https://www.cosmos.esa.int/web/gaia/ dpac/consortium)

\section*{REFERENCES}


\newpage

\end{document}